\documentclass[aps,prc,nofootinbib,reprint]{revtex4-1}

\usepackage{blindtext}
\usepackage{amsfonts}
\usepackage{tensor}
\usepackage{graphicx}
\usepackage{pict2e}

\usepackage{amssymb, amsmath,environ}

\usepackage{color}

\newcommand{\blk}{\color{black}}


\newcommand{\bse}{\begin{subequations}}
	\newcommand{\ese}{\end{subequations}}
\newcommand{\be}{\begin{equation}}
\newcommand{\ee}{\end{equation}}

\NewEnviron{eqsplit}{%
	\begin{equation}\begin{split}
	\BODY
	\end{split}\end{equation}
}

\newcommand{\bea}{\begin{eqnarray}}
\newcommand{\eea}{\end{eqnarray}}
\newcommand{\ba}{\begin{array}}
	\newcommand{\ea}{\end{array}}

\newcommand{\nn}{{\nonumber}}
\newcommand{\K}{\mathcal{K}}
\newcommand{\Q}{\mathcal{Q}}
\newcommand{\A}{\mathcal{A}}
\newcommand{\PP}{\mathcal{P}}

\newcommand{\E}{\mathcal{E}}
\newcommand{\V}{\mathcal{V}}

\newcommand{\la}{\langle}
\newcommand{\ra}{\rangle}
\newcommand{\ve}{\varepsilon}

\begin{document}

\title{Standardized Cumulants of Flow Harmonic Fluctuations}

\author{Navid Abbasi$^{1}$}
\email[]{Abbasi@ipm.ir}

\author{Davood Allahbakhshi$^{1}$}
\email[]{Allahbakhshi@ipm.ir}

\author{Ali Davody$^{1,2}$}
\email[]{Davody@ipm.ir}

\author{Seyed Farid Taghavi$^{1}$}
\email[]{s.f.taghavi@ipm.ir}

\affiliation{$^{1}$School of Particles and Accelerators, Institute for Research in Fundamental Sciences (IPM), P.O. Box 19395-5531, Tehran, Iran}
\affiliation{$^{2}$Institute of Theoretical Physics, Regensburg University, 93040 Regensburg,	Germany}

\begin{abstract}
The distribution of flow harmonics in heavy ion experiment can be characterized by standardized cumulants. 
 We first model the ellipticity and power parameters of the elliptic-power distribution by employing MC-Glauber model.
Then we use the elliptic-power distribution together with the hydrodynamic linear response approximation to study the two dimensional standardized cumulants of elliptic and triangular flow ($v_2$ and $v_3$) distribution.  For the second harmonic, it turns out that finding two dimensional cumulants in terms of $2q$-particle correlation functions $c_2\{2q\}$ is limited to the skewness. We also show that $c_3\{2\}$, $c_3\{4\}$, and $c_3\{6\}$, are related to the second, fourth, and sixth standardized cumulants of the $v_3$ distribution, respectively. The cumulant $c_{n}\{2q\}$ can be also written in terms of $v_n\{2q\}$. Specifically, $-(v_3\{4\}/v_3\{2\})^4$ turns out to be the kurtosis of the $v_3$ event-by-event fluctuation distribution. We introduce a new  parametrization for the distribution $p(v_3)$ with $v_3\{2\}$, kurtosis and sixth-order standardized cumulant being its free parameters.  Compared to the Gaussian distribution, it indicates a more accurate fit with experimental results. Finally, we compare the kurtosis obtained from simulation with that of extracted from experimental data for the $v_3$ distribution.

\end{abstract}

\maketitle

\section{Introduction}

There is a strong belief that the matter produced in the heavy ion collision experiments in the both Relativistic Heavy Ion Collider (RHIC) and the Large Hadron Collider (LHC) has a collective behavior. This is experimentally confirmed by measuring the second Fourier harmonic of the particle momentum azimuthal distribution, namely the \textit{elliptic flow}, $v_2$  \cite{Ackermann:2000tr,Aamodt:2010pa}. In fact, the almond shape of the initial energy density in noncentral collisions manifests itself in $v_2$.  Moreover, there exist other \textit{flow harmonics} (Fourier harmonics) such as triangle flow, $v_3$ \cite{Alver:2010gr}, which corresponds to the event-by-event position fluctuations of nucleons inside the nucleus. The triangular flow as well as other flow harmonics have been observed in RHIC \cite{Adare:2011tg,Sorensen:2011fb} and LHC \cite{ALICE:2011ab,Aamodt:2011by,Li:2011mp,Aad:2014vba}.

The reaction plane angle in a single collision is not an accurate observable in the experiments. If we had prior knowledge about the reaction plane, then we would obtain different values for each flow harmonic of the events in the same centrality class. However, one would still be able to extract the flow harmonics of many events in the same centrality class with even unknown reaction plane angle. One way to do that is to use the multiparticle azimuthal correlation function, $c_n\{2k\}$ \cite{Borghini:2000sa,Borghini:2001vi}. 

In Reference\cite{Yan:2014afa}, it is shown that the fluctuations of initial anisotropy $\ve_n$, generated by different initial condition Monte Carlo generators, can be described by elliptic-power distribution. This distribution is not exact Gaussian. As a result, after the hydrodynamic evolution, we expect that the $v_n$ distribution of an ensemble of events in the same centrality class not to be exactly Gaussian, too. There are certain statistical quantities such as skewness, kurtosis, etc, which quantify the deviation of a given distribution from Gaussianity. It has been shown that the fine splitting between $v_2\{4\}$ and $v_2\{6\}$ is the consequence of nonzero skewness in $v_2$ distribution \cite{Giacalone:2016eyu}.

In this work, to study the distribution of $v_2$ we first use a simple model of heavy-ion collision, the elliptic-power distribution together with the linear response of hydrodynamics. It turns out that by considering $c_2\{2\}$ to $c_2\{8\}$, the only quantity which can be experimentally extracted from $v_2$ distribution would be the skewness. We also show that for $v_3$ distribution, both ratios $-(v_3\{4\}/v_3\{2\})^4$ and $4(v_3\{6\}/v_3\{2\})^6$ are indicating the deviation of $v_3$ distribution from Gaussianity. Similar quantities have been studied before in Refs.~\cite{Bhalerao:2011bp,Giacalone:2017uqx}; however, here, we find their relation with standardized cumulants as well. In addition, we introduce  a new parametrization for the distribution function $p(v_3)$ [see \eqref{aDistribution}]  which has a small deviation  from Gaussianity, identified by two standardized cumulants. These cumulants can be found by fitting the $v_3$ distribution with experimental data. Finally, in a more realistic model, we use the iEBE-VISHNU event generator \cite{Shen:2014vra} and compare the kurtosis of $v_3$, found by an event generator with that of obtained from experimental data.

\section{elliptic-power Distribution}\label{initialSection}

In this section, we will review a simple and interesting model of the heavy-ion collision initial state, introduced in References \cite{Danielewicz:1983we,Ollitrault:1992bk,Yan:2014afa}. Consider $N$ independent point-like sources distributed in a two-dimensional (2D) plane with a 2D Gaussian probability distribution. This distribution function could have different widths along two directions. We can imagine the sources as the location of nucleon-nucleon collision in the Glauber model. Using this distribution, as we will see in the following, one can find a distribution for the  initial anisotropy $\ve_n$ which is called the elliptic-power distribution. Although this model is more simple than the MC-Glauber \cite{Broniowski:2007nz,Alver:2008aq,Loizides:2014vua}, MC-KLN \cite{Kharzeev:2001yq,Kharzeev:2004if} and IP-Glasma \cite{Schenke:2012wb} models, it has been shown in Ref.~\cite{Yan:2014afa} that the elliptic-power distribution fits perfectly with the $\ve_n$ distribution generated by the more complex models.

 \begin{figure}
 	\begin{center}
 		\begin{tabular}{c}
 			\includegraphics[scale=0.4]{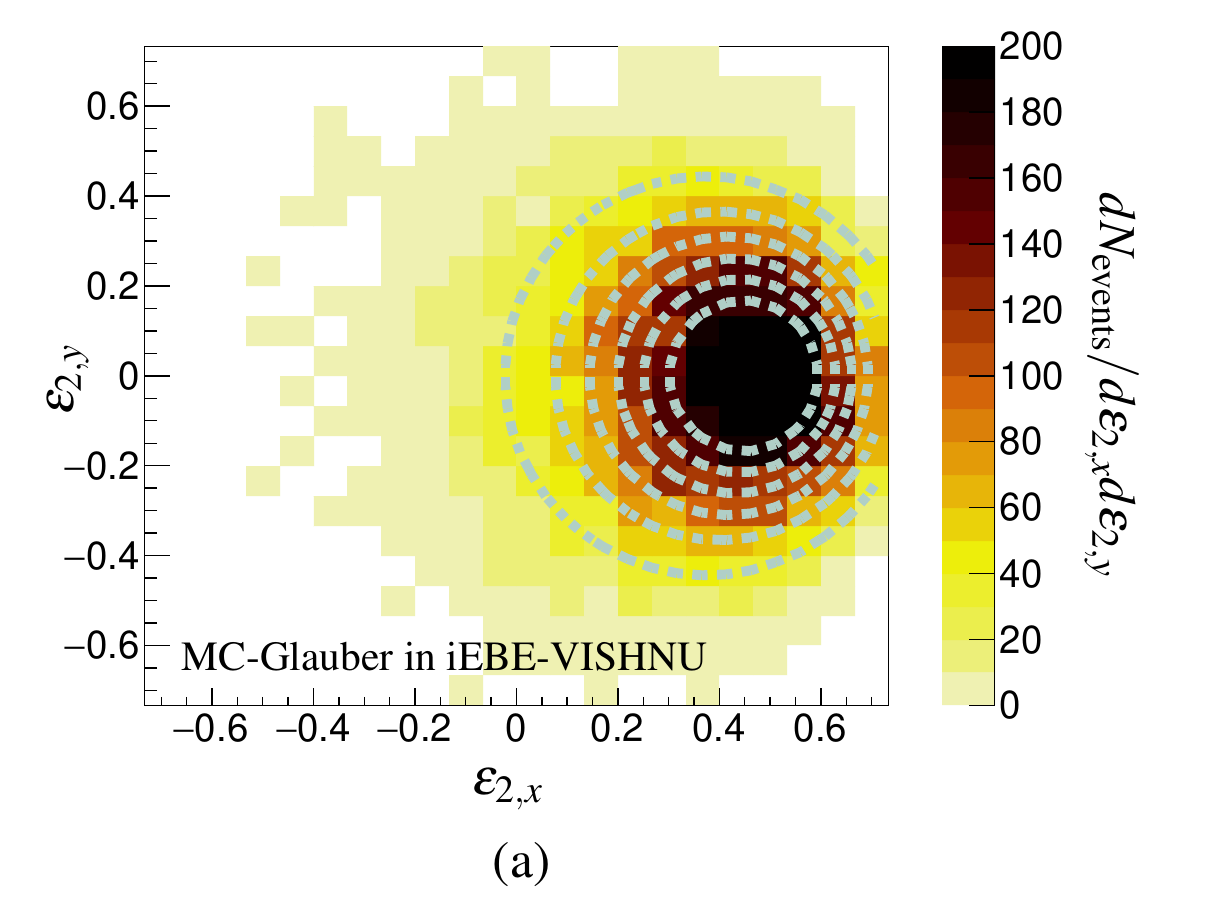}			\vspace*{0.4cm}\\
 			\includegraphics[scale=0.4]{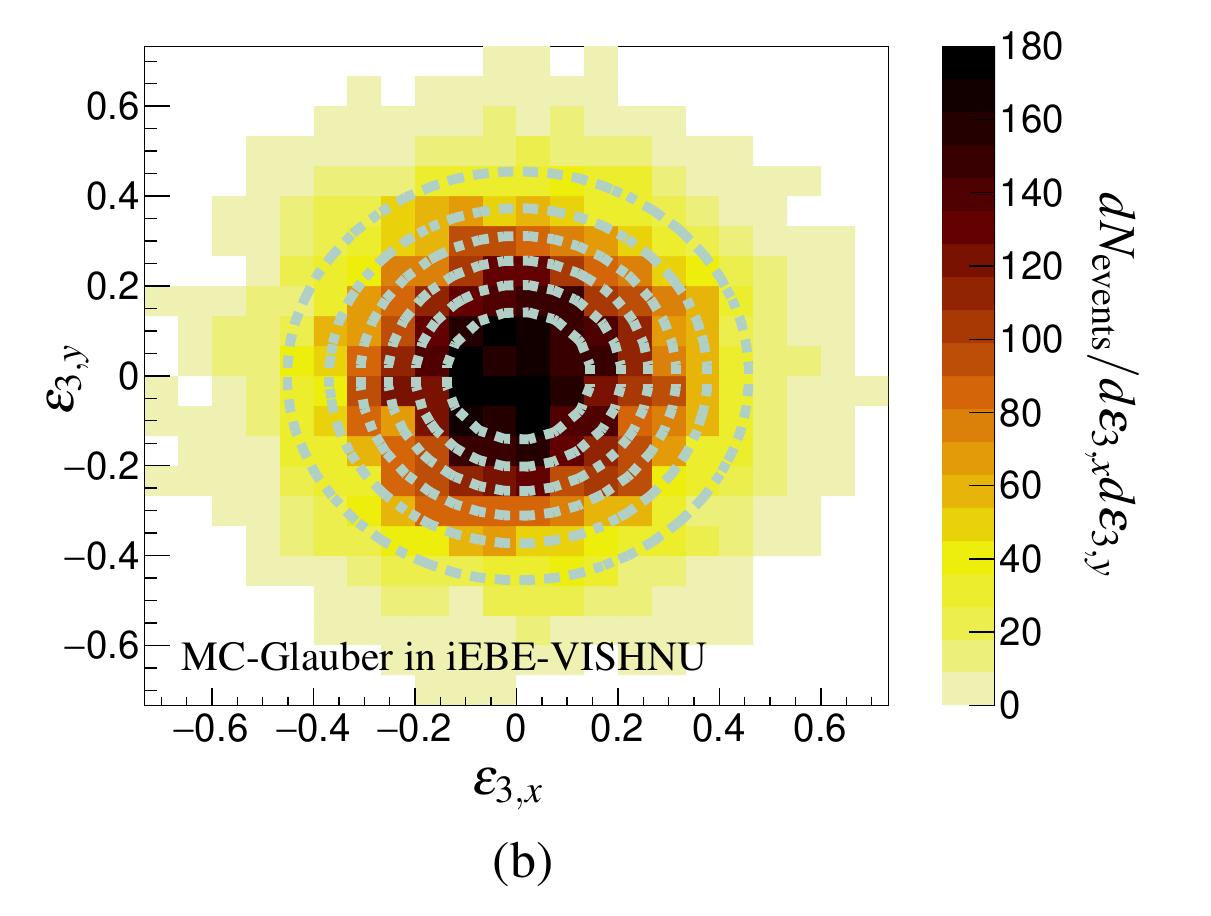}\vspace*{0.3cm}\\

 		\end{tabular}	
 		\caption{(Color online) The eccentricity and triangularity distribution of 14\,000 events for MC-Glauber model in $50-55\%$ centrality class of Pb-Pb collision, generated by iEBE-VISHNU (yellow spectrum). The elliptic-power distribution \eqref{EPD} is indicated by light-blue dashed contours. The ellipticity and power are obtained by fitting: (a) $n=2$, $\alpha\simeq8.70$, and $\varepsilon_0\simeq0.40$; (b) $n=3$, $\alpha=9.54$, and $\varepsilon_0 = 0.00$.} 
 		\label{EllipticPowerFit}
 	\end{center}
 \end{figure}

   In this model, the energy density is given by
\bea\label{discreteDist}
\rho(X,Y)=\rho_0\sum_{i=1}^{N}\delta(X-X_i)\delta(Y-Y_i)
\eea
where $(X_i,Y_i)$ is the position of $i^{th}$ source on the plane. In order to quantify the shape of each randomly generated event, we use the 2D Fourier analysis of $\rho(X,Y)$ developed in Ref.~\cite{Teaney:2010vd}. Introducing the averaging over energy density of a single event 
$$\{\cdots\}=\frac{\int \cdots \rho(X,Y) dXdY}{\int \rho(X,Y) dXdY}, $$
 we define the complex quantity $\boldsymbol{\ve}_n $ as
\begin{eqsplit}\label{eccentisities}
    \boldsymbol{\ve}_n  =\ve_n e^{in\Phi_n} \equiv\frac{\{r^n e^{in\varphi}\}}{\{r^n\}},
\end{eqsplit} 
where $r$ and $\varphi$ are radial and azimuthal coordinates in the $X$-$Y$ plane. For $n=1$ in this relation, we have to replace $r$ with $r^3$ \cite{Teaney:2010vd}.\footnote{ The other method to characterize the initial conditions is using the Bessel-Fourier modes \cite{Floerchinger:2013rya}. The advantage of using these modes is that they are making a complete basis and the fluid dynamic evolution can be studied for each mode separately. }   Occasionally, we use the Cartesian notation wherein $\varepsilon_{n,x}=\ve_n\cos n\Phi_n$ and $\varepsilon_{n,y}=\ve_n\sin n\Phi_n$. Using \eqref{discreteDist}, we can specifically find $\varepsilon_{2,x}$ and $\varepsilon_{2,y}$ as follows,
\begin{eqsplit}\label{eccentrisities}
	\varepsilon_{2,x}=\frac{\sum_{i=1}^{N}(X_i^2-Y_i^2)}{\sum_{i=1}^{N}(X_i^2+Y_i^2)},\quad \varepsilon_{2,y}=\frac{2\sum_{i=1}^{N} X_i Y_i}{\sum_{i=1}^{N}(X_i^2+Y_i^2)}.
\end{eqsplit}
The above $\ve_{2,x}$ is indicating how much the randomly generated event is almond shaped while, $\ve_{2,y}$ shows how much the almond is rotated in the $X$-$Y$ plane. In Refs.~\cite{Danielewicz:1983we,Ollitrault:1992bk}, it has been shown that if we randomly generate several events with a specific width of Gaussian distribution, then the probability distribution of events with respect to $\varepsilon_{2,x}$ and $\varepsilon_{2,y}$ is given by
\bea\label{EPD}
p(\varepsilon_{n,x},\varepsilon_{n,y})=\frac{\alpha}{\pi}(1-\varepsilon_0^2)^{\alpha+1/2}\frac{(1-\varepsilon_{n,x}^2-\varepsilon_{n,y}^2)^{\alpha-1}}{(1-\varepsilon_0\varepsilon_{n,x})^{2\alpha+1}}.\nn\\
\eea
In the above, we follow  Ref.~\cite{Yan:2014afa} and  use $\ve_n$ not only  for $n=2$, but also for $n>2$. This relation is called \textit{elliptic-power} distribution. In this distribution, the  \textit{ellipticity} $\ve_0$ and \textit{power} $\alpha=(N-1)/2$ are two unknown free parameters.
Note that for $\alpha\gg1$, this distribution reduces to a 2D Gaussian distribution.

In Ref.~\cite{Yan:2014afa}, the parameters $\varepsilon_0$ and $\alpha$ are obtained by fitting the function \eqref{EPD} with the azimuthally integrated distributions generated by different models. As we expect, the result depends on the model we are studying and also on the value of $n$. It is worth mentioning that for $\varepsilon_{3}$ the best fit is obtained by setting $\varepsilon_0=0$, because we do not expect any average value for this parameter. Specifically, if we set $\ve_0=0$ and integrate over $\varphi$, then we find the \textit{power} distribution \cite{Yan:2013laa}
\bea\label{powerDist}
p(\ve_n)= 2\alpha \ve_n(1-\ve_n^2)^{\alpha-1}.
\eea

Here we calculate the $\ve_0$ and $\alpha$ without integrating over the azimuthal angle in \eqref{EPD}. We generate up to 14\,000 initial states of Pb-Pb collision with center-of-mass energy $\sqrt{s}=2.76\,$TeV using the MC-Glauber model implemented in the iEBE-VISHNU generator \cite{Shen:2014vra}.\footnote{ The initial eccentricities $\ve_{n,x}$ and $\ve_{n,y}$ are calculated from initial state energy density in iEBE-VISHNU. } To find the $\varepsilon_0$ and $\alpha$, we fit \eqref{EPD} with the distribution found by filling a 2D histogram of $\varepsilon_{n,x}$ and $\varepsilon_{n,y}$. In Fig.~\ref{EllipticPowerFit}, we have depicted the histogram and elliptic-power distribution for $50-55\%$ centrality class. The result  of fitting  for harmonic $n=2$ is $\alpha=8.699\pm 0.076$, $\varepsilon_0=0.400\pm 0.002$, while for $n=3$ is $\alpha=9.543\pm 0.083$, $\varepsilon_0 = 0.004\pm 0.002$. 

 One can do the same calculation for different centralities and find $\alpha$ and $\varepsilon_0$. The result is presented in Fig.~\ref{alpha_epsilon_central}. As can be seen in  Fig.~\ref{alpha_epsilon_central}(a), by increasing the centrality, $\alpha$ decreases, in agreement with Ref.~\cite{Yan:2014afa}.  From Fig.~\ref{alpha_epsilon_central}(b), one finds that for $n=2$ the parameter $\varepsilon_0$ is nonzero for noncentral collisions. The reason is that the net ellipticity is nonzero in noncentral collisions due to the collision geometry. However, for $n=3$, the parameter $\varepsilon_0$ is almost zero for all centralities because there is no net triangularity for spherical ions and symmetrical collisions. The numerical values of $\alpha$ and $\varepsilon_0$ will be used in the following sections.

For $n=2$, it is already well known \cite{Yan:2014afa,Giacalone:2016eyu} that the distribution is left skewed in the $\varepsilon_{2,x}$ direction. We have demonstrated this result in a two-dimensional histogram in Fig.~\ref{EllipticPowerFit}(a). For $n=3$, however, $\varepsilon_0$ is almost zero [Fig.~\ref{alpha_epsilon_central}(b)] and in this case, no apparent skewness\footnote{We define the skewness systematically in Sec.~\ref{cumulSec}. } is observed in the distribution.  The skewness of the elliptic-power distribution can be explained in the following. Suppose the sources are distributed via a 2D Gaussian with the width in the $x$ axis ($\sigma_x$) being larger than that in the $y$ axis ($\sigma_y$) and also with the vanishing cross term. The latter means the larger axis of the almond is fixed along the $x$ axis. Using these assumptions, it turns out that the average of $\varepsilon_{2,y}$ is equal to zero while $\varepsilon_{2,x}$ gets a nonzero average. Based on the above assumptions, no skewness would be observed in the $y$ direction while the distribution is skewed in the $x$ direction. The reason for the latter statement is as follows. The distribution along the $y$ axis is narrower than that of along the $x$ axis. In other words, the sources are more probable to be generated along the $x$ axis rather than the $y$ axis. Therefore, the distribution of the $\varepsilon_{2,x}$ is more concentrated on the right side of the average, which means that it is left-skewed. For central collisions [or for $n=3$ in Fig.~\ref{EllipticPowerFit}(b)] with $\la\ve_{2,x}\ra=0$ the distribution in Fig.~\ref{EllipticPowerFit}(a) becomes rotationally symmetric and consequently nonskewed.\footnote{ In Ref.~\cite{Giacalone:2016eyu},  the skewness of the distribution $p(\ve_{2,x},\ve_{2,y})$ is explained as follows: By construction  we have $\ve_2=\sqrt{\ve_{2,x}^2+\ve_{2,y}^2}\leq1$  [see \eqref{eccentisities}]. As a result, $p(\ve_{2,x},\ve_{2,y})$ is bounded in the  unit circle, and $\la \ve_{2,x} \ra \neq 0$ leads to the skewness.   }

\section{Cumulant Analysis of Probability Distribution}\label{cumulSec}

In the previous section, we observed that the elliptic-power distribution is skewed for $\varepsilon_0\neq 0$ and the same feature was observed for the MC-Glauber. In order to study the distribution of $\boldsymbol{\ve}_n$ (and flow harmonics) we employ the cumulant analysis. We first review the terminology used in the present work and then find the explicit form of standardized cumulants (will be defined shortly) of the elliptic-power distribution. 

\subsection{Cumulants: Review and terminology}\label{terminologu}

\begin{figure}[t!]
	\begin{center}
		\begin{tabular}{c}
			\includegraphics[scale=0.4]{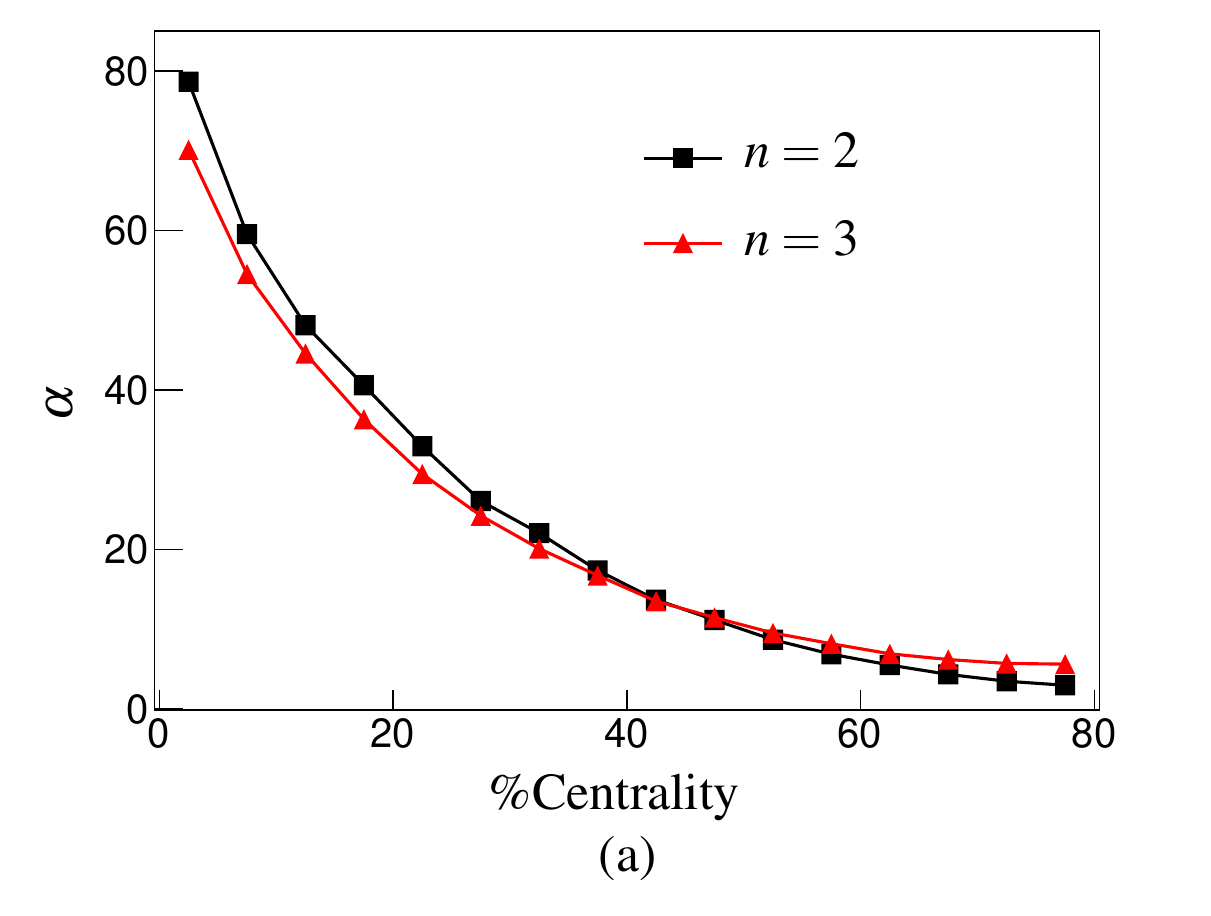}			\vspace*{1.0cm}\\
			\includegraphics[scale=0.4]{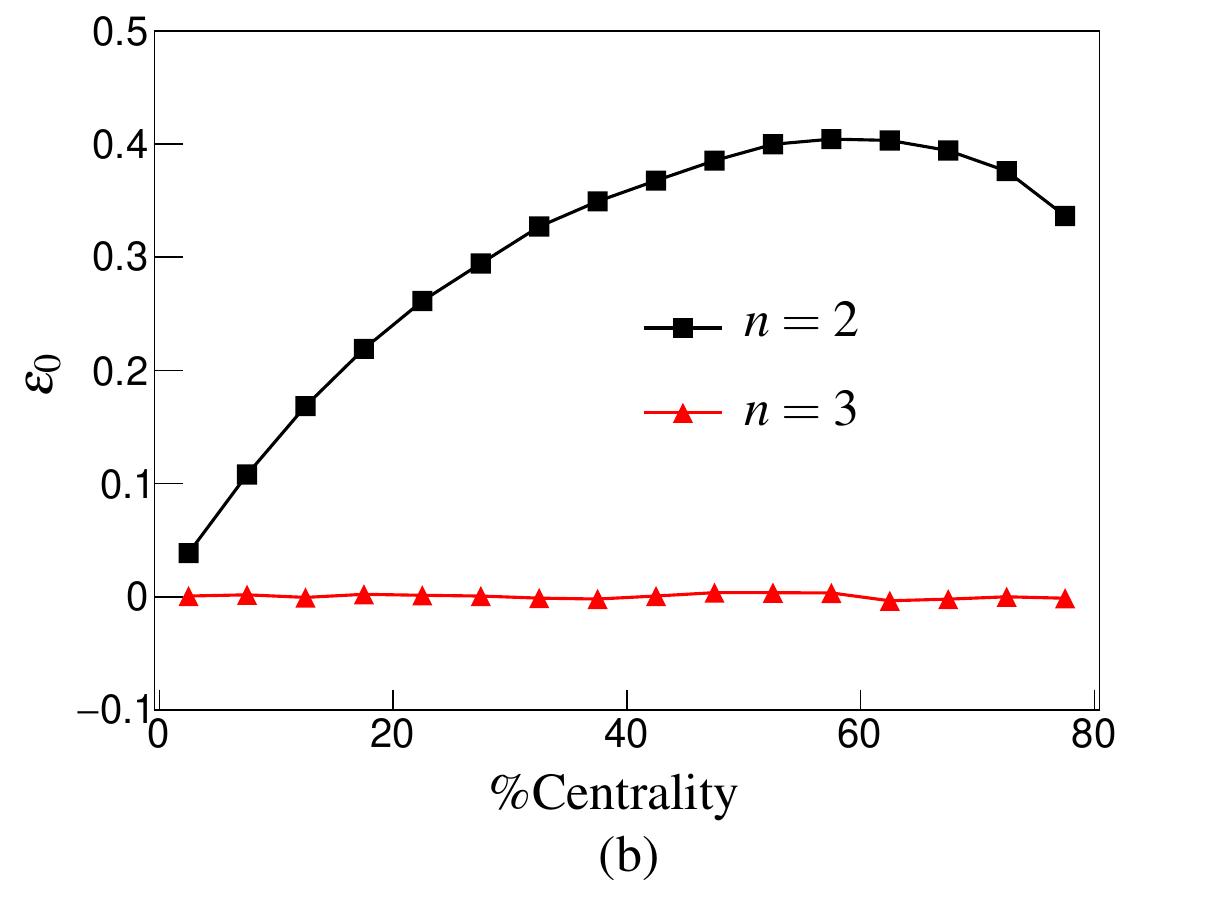}					
		\end{tabular}
		\caption{(Color online) The power (a)  and ellipticity (b) obtained by fitting \eqref{EPD} with MC-Glauber initial states for harmonics $n=2,3$.  The error bars are smaller than the size of the points. } 
		\label{alpha_epsilon_central}
	\end{center}
\end{figure}

The cumulants of a distribution $\PP(\xi)$ are obtained from the generating function $\log \la e^{\lambda \xi} \ra$. If we expand this function around $\lambda=0$, then the cumulant $\kappa_n$ will be the coefficient of the $\lambda^n/n!$. The advantages of using cumulants instead of moments are that they are homogeneous and shift invariant (except $\kappa_1$) and $\kappa_{n\geq 3}=0$ for the normal distribution.

In statistics, the standardized central moments $\gamma_1=\la (\xi-\la \xi\ra)^3\ra /\sigma^3$ and $K=\la (\xi-\la \xi\ra)^4\ra /\sigma^4$ are called skewness and kurtosis respectively. Recalling $\kappa_2\equiv \sigma^2$, one can simply see that $\gamma_1=\kappa_3/\kappa_2^{3/2}$ and  $K=\kappa_4/\kappa_2^2+3$. According to the properties of the cumulants, the kurtosis of a Gaussian distribution is equal to 3. For this reason, it is common to call $\gamma_2=K-3$ as the kurtosis. We will use the latter terminology in this work. In general, we define the standardized cumulants of a distribution as follows,
\bea\label{skewGen1D}
\gamma_{q-2}=\frac{\kappa_q}{\kappa_2^{q/2}}.
\eea
In most part of this paper, we deal with two-dimensional distributions and therefore need to use the 2D (standardized) cumulants. Similarly to the 1D case, we can find the cumulants by expanding the following generating function,
\bea\label{cumulGen}
\log\la e^{\lambda_x \xi_x+\lambda_y \xi_y}\ra =\sum_{m,n=0}\frac{\lambda_x^m \lambda_y^n}{m! n!} \A_{mn},\quad
\eea
where $(\xi_x,\xi_y)$ is a 2D random variable with a 2D distribution function $\PP(\xi_x,\xi_y)$.\footnote{In this manuscript, we refer to $\A_{mn}$ as the \textit{cumulant} and to the $c_n\{2k\}$ as $2k$-particle correlation function.} From \eqref{cumulGen}, $\A_{mn}$ is found in terms of the moments $\la\xi_x^{p}\xi_y^{q}\ra$.\footnote{In Ref.~\cite{Giacalone:2016eyu}, $\A_{30}$ and $\A_{12}$ are shown by $s_1$ and $s_2$, respectively.} In the following, we consider $m+n$ as the order of the $\A_{mn}$ cumulant.  It is worth mentioning that the cumulants of a normal distribution  with order higher than two are equal to zero. Also it can be shown that the cumulant statistical error of a sample with $N$ entries is proportional to $\frac{1}{N}.$\footnote{The method of finding the explicit form of the errors can be found in the statistic textbooks such as Ref.~\cite{kendallBook}. }

In order to generalize the notion of skewness, kurtosis, etc. into 2D dimensions, we can simply replace \eqref{skewGen1D} with the following expression:
\begin{eqsplit}\label{normalCumul}
	\hat{\A}_{mn}=\frac{\A_{mn}}{\sqrt{\A_{20}^m\A_{02}^n}},
\end{eqsplit}
where clearly we have $\hat{\A}_{20}=\hat{\A}_{02}=1$. In the following, we call $\hat{\A}_{mn}$ as (2D) standardized cumulants.

\subsection{Moments and cumulants of elliptic-power distribution}

Now we specifically concentrate on the cumulants of the elliptic-power distribution. We show the cumulant obtained from $\boldsymbol{\ve}_n$ distribution by $\E_{kl}^{(n)}$. In order to find $\E_{kl}^{(n)}$, we first have to compute the moments of distribution:
\bea\label{moments}
\la\ve^{k}_{n,x}\ve^{l}_{n,y}\ra=\int d\ve_{n,x}d\ve_{n,y}\,\ve^{k}_{n,x}\ve^{l}_{n,y}\,p(\ve_{n,x},\ve_{n,y}).
\eea
By considering the symmetries of \eqref{EPD}, some of the moments identically vanish. Let us recall that for $n=2$ the $\ve_0$ is nonzero for noncentral collisions, and, hence, the probability $p(\ve_{2,x},\ve_{2,y})$ is not symmetric under $\ve_{2,x}\to -\ve_{2,x}$ in this case. However, $p(\ve_{2,x},\ve_{2,y})$ is an even function with respect to parameter $\ve_{2,y}$ which immediately leads to $\la \ve^{k}_{2,x}\ve^{2l+1}_{2,y}\ra=0$. Under the above considerations, if we use the explicit form of the cumulants by extracting them from \eqref{cumulGen}, we find $\E^{(2)}_{21}=\E^{(2)}_{03}=0$.

\begin{figure*}[t!]
	\begin{center}
		\begin{tabular}{cc}
			\includegraphics[scale=0.4]{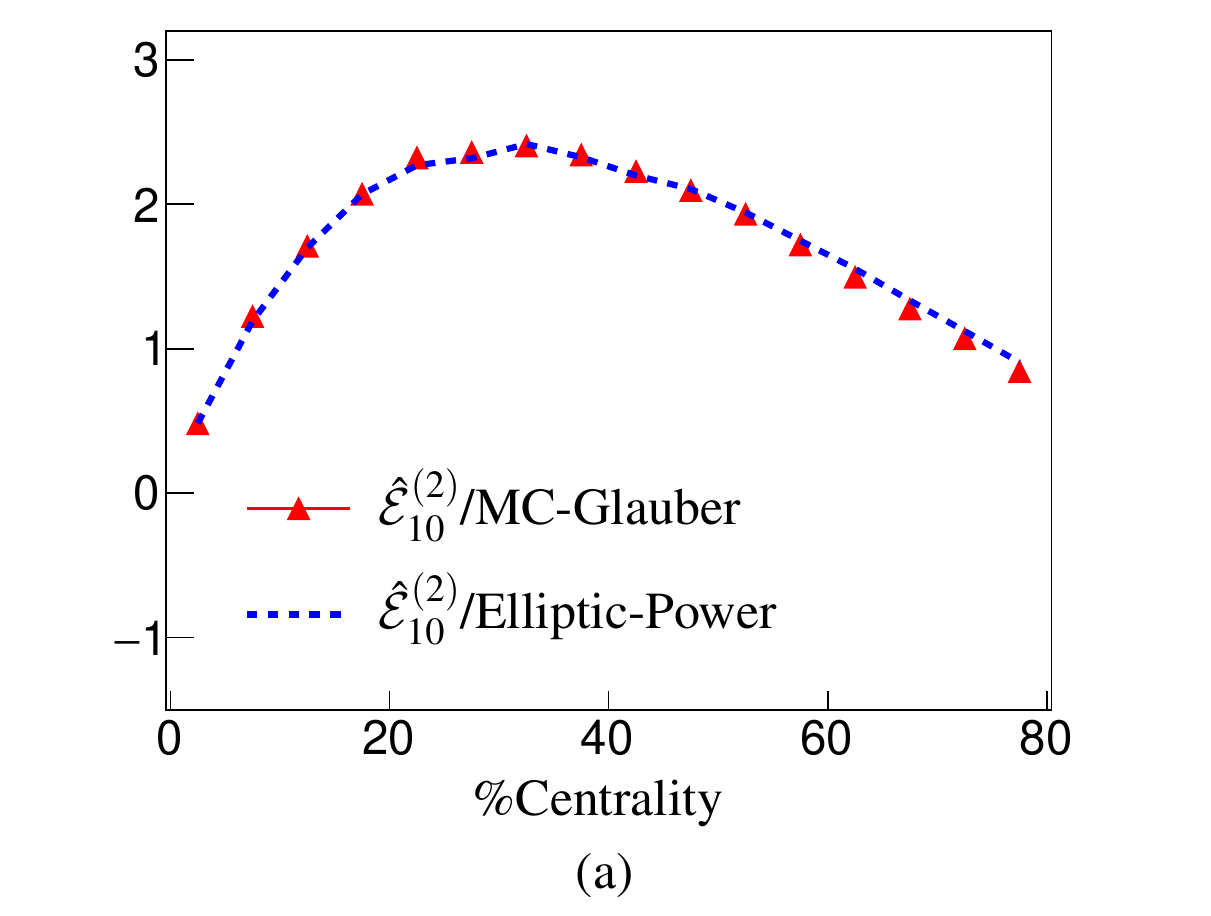}	
			& 			\includegraphics[scale=0.4]{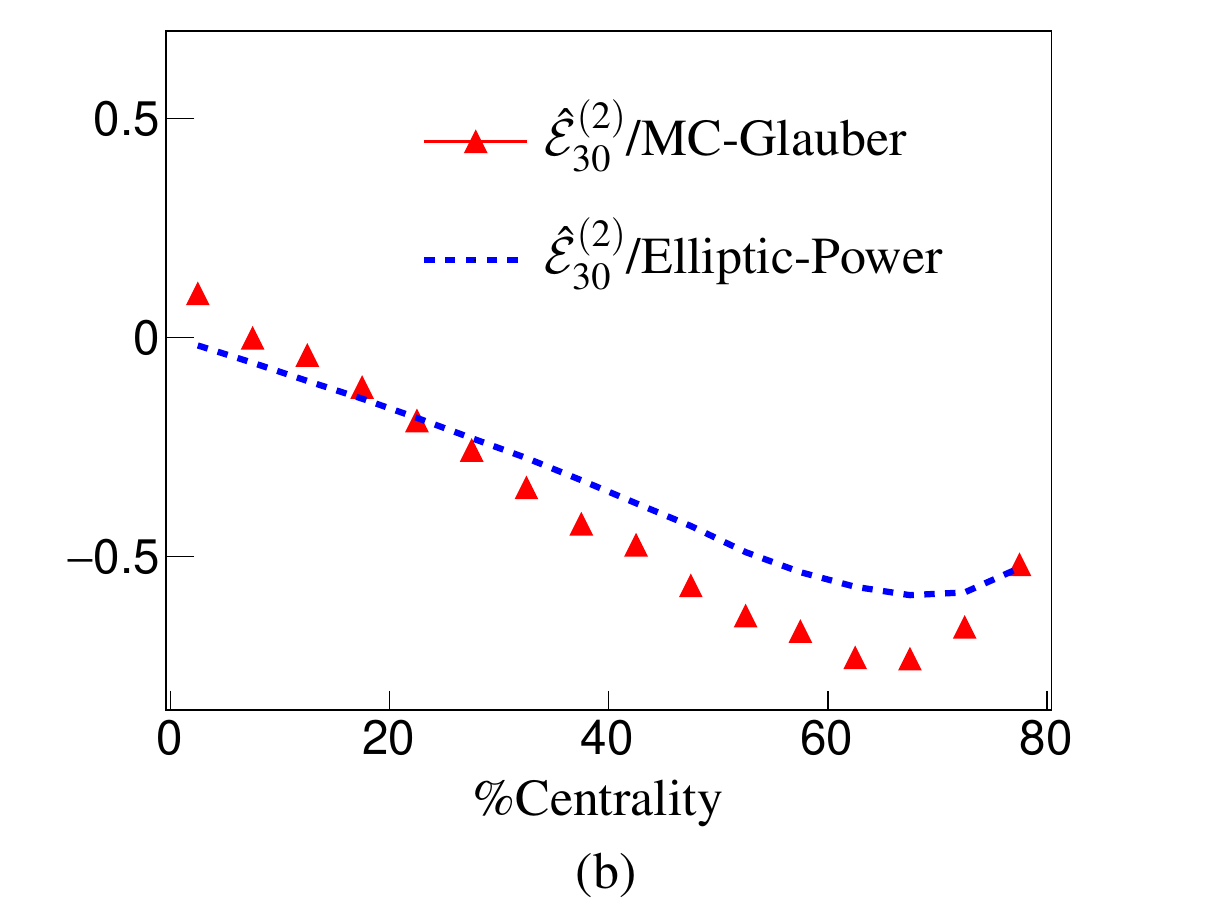}
			\\		
			\includegraphics[scale=0.4]{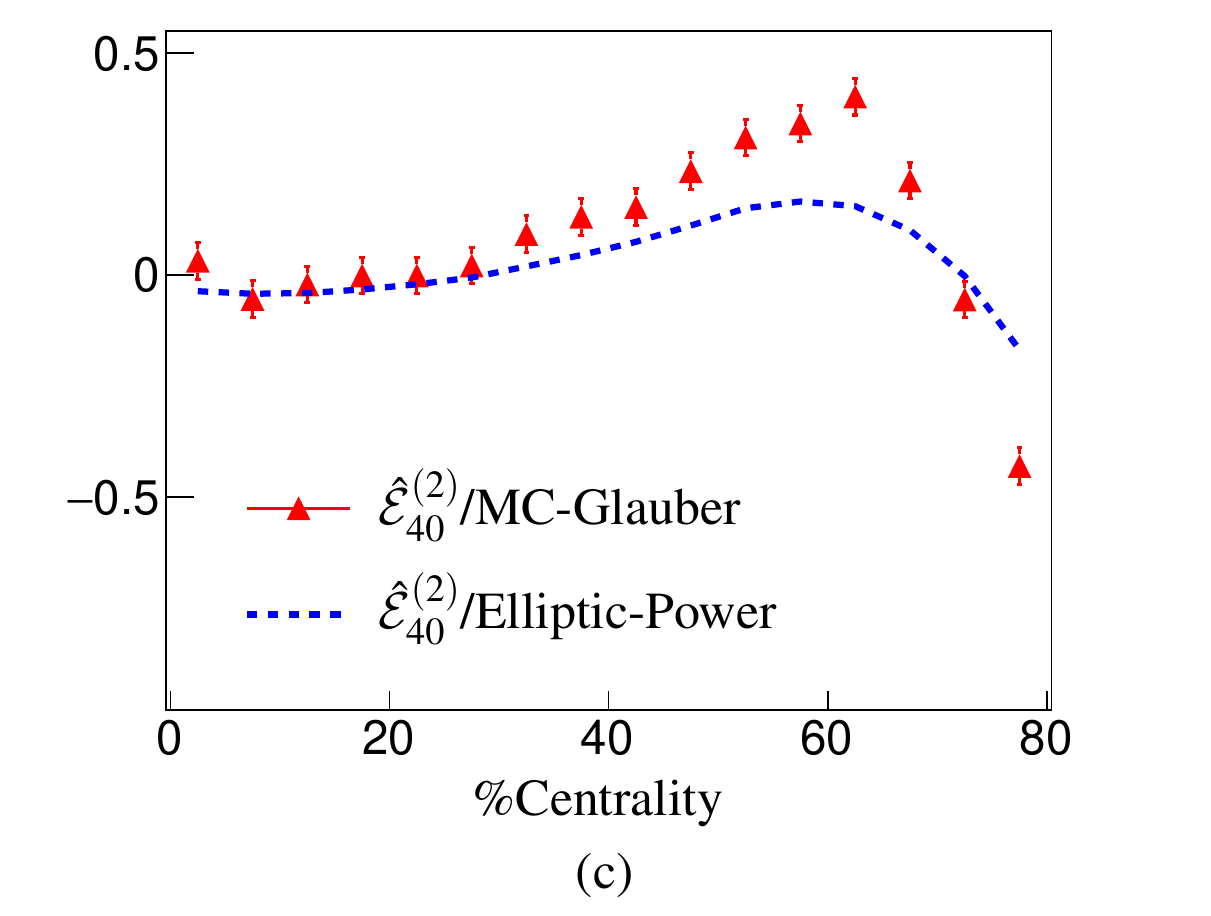}
			& 			\includegraphics[scale=0.4]{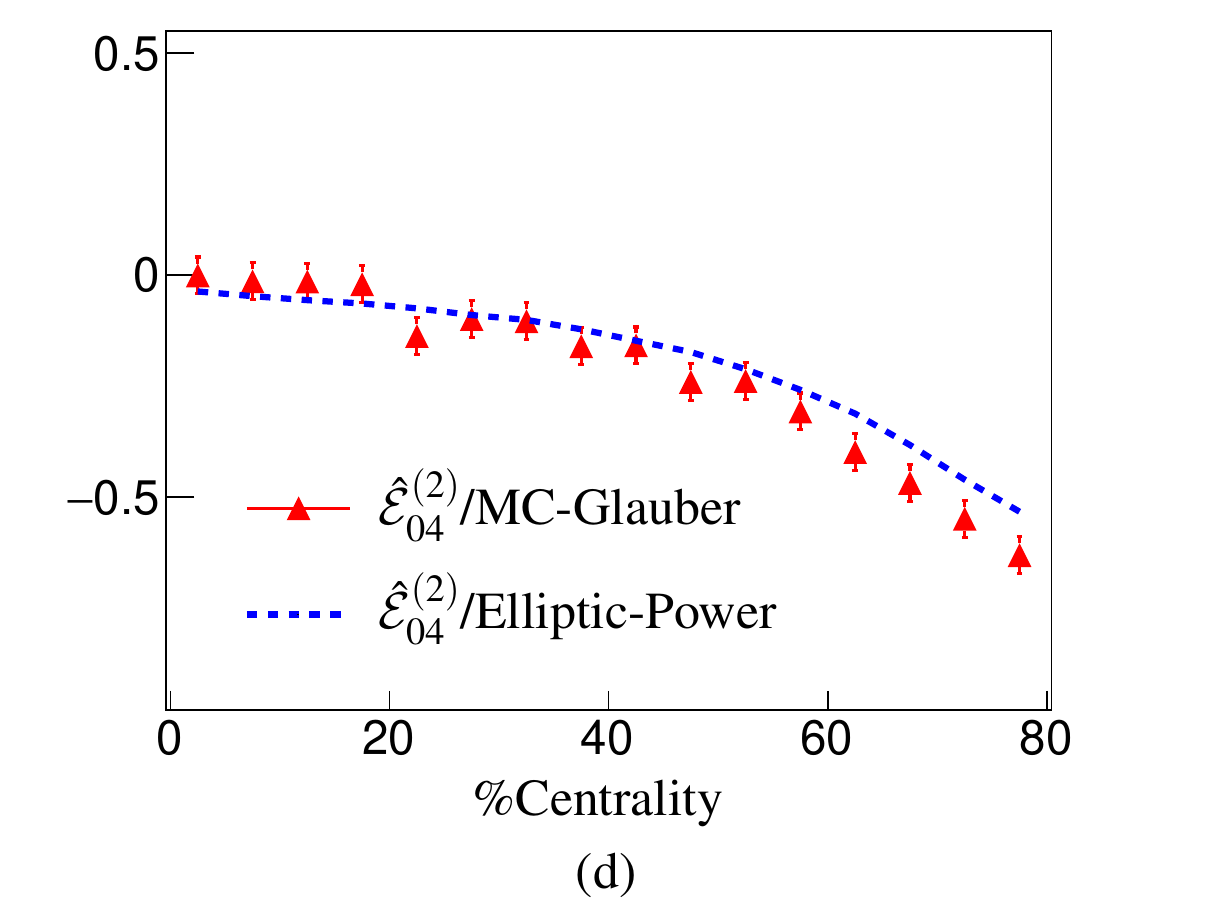}	
		\end{tabular}
		\caption{(Color online) Some nonzero standardized cumulants obtained from event-by-event fluctuation distribution. The blue dashed curve (red dots) is related to the standardized cumulants $\hat{\E}_{mn}$ obtained from the  initial anisotropy $\boldsymbol{\ve}_2$ distribution acquired from elliptic-power (MC-Glauber) distribution. The error bars indicate the statistical errors.} 
		\label{n2cumulants}
	\end{center}
\end{figure*}

On the other hand, for $n=3$ we have $\ve_0\simeq 0$, which means $p(\ve_{3,x},\ve_{3,y})$ is even with respect to both parameters $\ve_{3,x}$ and $\ve_{3,y}$. Consequently, the only nonzero moments are $\la \ve^{2k}_{3,x}\ve^{2l}_{3,y}\ra$. In other words, for $n=3$, all odd order cumulants are  equal to zero, i.e., $\E^{(3)}_{kl}=0$ for $k+l=2q+1$. As a result, the nonzero and nontrivial standardized cumulants appear from the fourth order. For $\boldsymbol{\ve}_3$, the other observation from \eqref{EPD} is that it is symmetric with respect to $\ve_{3,x}\leftrightarrow\ve_{3,y}$ which means $\E^{(3)}_{kl}=\E^{(3)}_{lk}$.

\begin{figure*}[htbp]
	\begin{center}
		\begin{tabular}{cc}
			\includegraphics[scale=0.4]{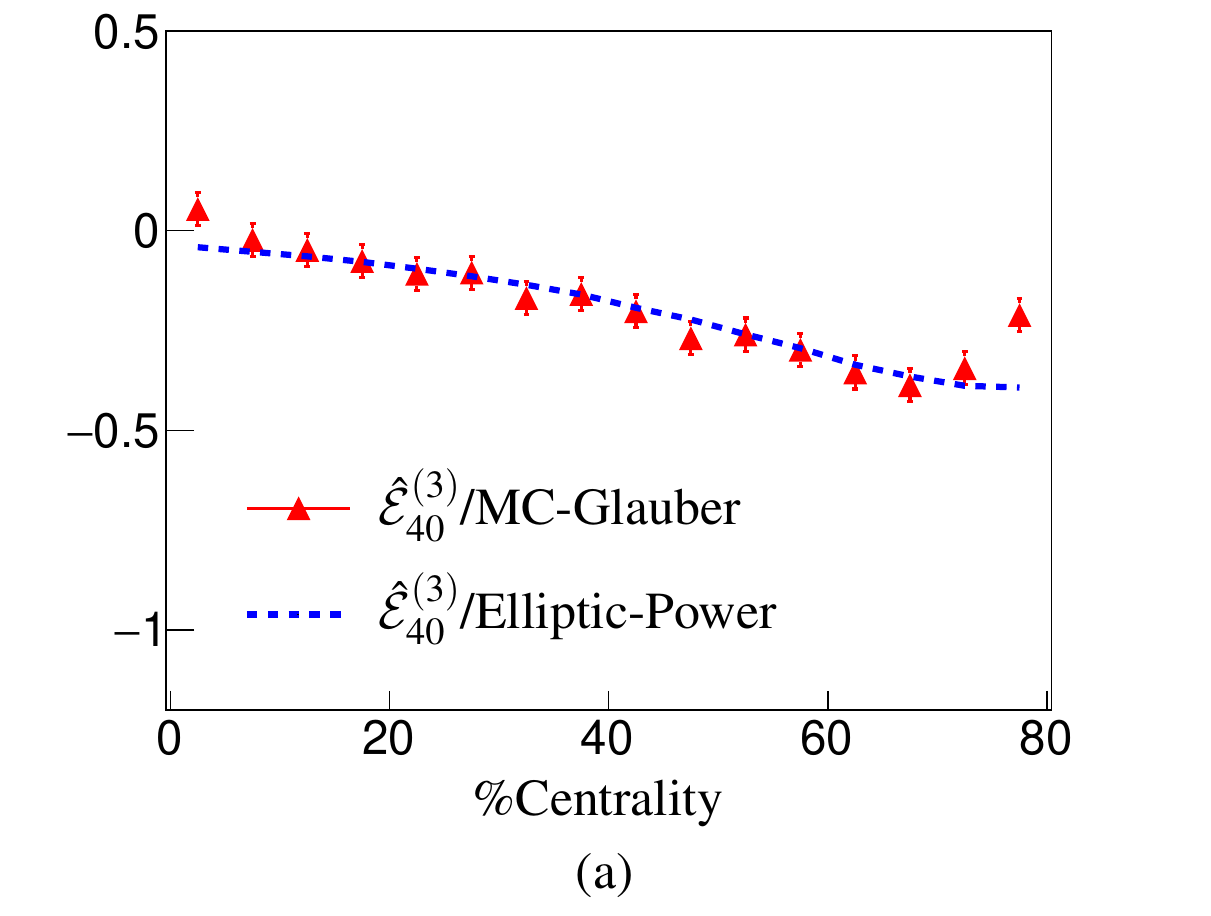}
				&			
 			\includegraphics[scale=0.4]{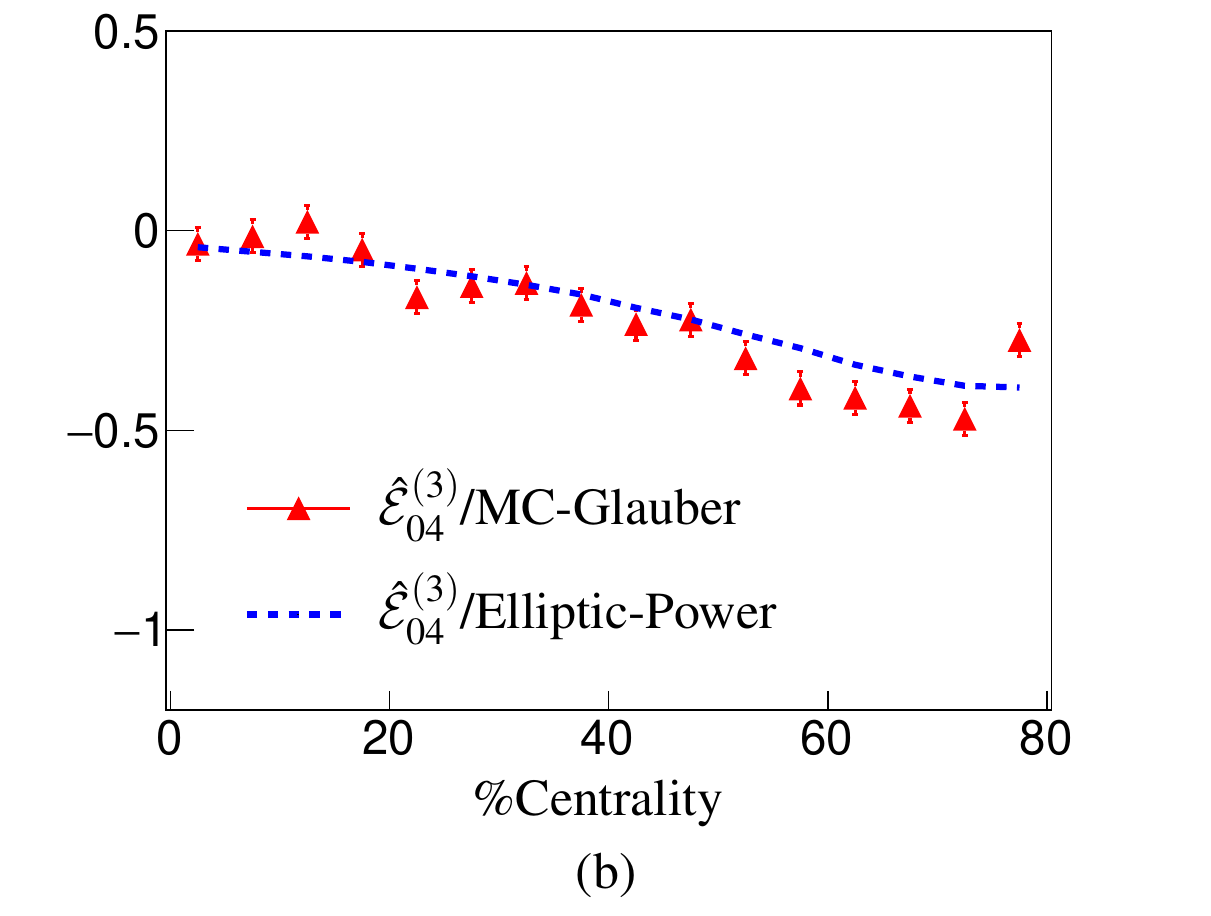}
 		\end{tabular}
		\caption{(Color online) Similar to Fig.~\ref{n2cumulants} for $n=3$.} 
		\label{n3cumulants}
	\end{center}
\end{figure*}

In general, we can find the moments of the elliptic-power distribution analytically. Introducing the following integral:
\bea\label{I_integral}
I_m(q,\alpha,\beta)=\int_{-1}^{1}dx\,\frac{x^m(1-x^2)^{\alpha}}{(1-q x)^{\beta}},
\eea
we are able to write the moments \eqref{moments} as follows:
\begin{eqsplit}
	 \la\ve^{k}_{n,x}&\ve^{l}_{n,y}\ra=\frac{\alpha}{\pi}(1-\varepsilon_0^2)^{\alpha+1/2}\\
	 &\times I_k(\varepsilon_0,\alpha+\frac{l-1}{2},2\alpha+1)\;I_l(0,\alpha-1,0).
\end{eqsplit}
The integral $I_m(q,\alpha,\beta)$ has analytical solution in terms of the hypergeometric functions (see Appendix \ref{integral}). Using this together with \eqref{cumulGen}, one can find cumulants of the elliptic-power distribution straightforwardly.

Let us now consider the implications of the above discussion to the case of heavy ion collisions. We know that in this case, both $\ve_0$ and $\alpha$ depend on the centrality (Fig.~\ref{alpha_epsilon_central}).  By knowing the centrality dependence of $\ve_0$ and $\alpha$,  we would obtain semi-analytical cumulants which can be used as a model for describing the heavy ion collision initial state.

\subsection{Cumulants: Elliptic-power vs. MC-Glauber}

 Having known the parameters $\alpha$ and $\ve_0$ from MC-Glauber (see Fig.~\ref{alpha_epsilon_central}),  we are able to compare the standardized cumulants $\hat{\E}^{(n)}_{kl}$ calculated from elliptic-power distribution with those of extracted from MC-Glauber. It is worth noting that we are \textit{modeling} the $\alpha$ and $\ve_0$ of the elliptic-power distribution by using MC-Glauber simulation. Therefore, we should keep in mind that some (not all) information of MC-Glauber distribution is already encoded in the elliptic-power (with modeled $\alpha$ and $\ve_0$) distribution. The results are depicted in Fig.~\ref{n2cumulants} for $n=2$ and Fig.~\ref{n3cumulants} for $n=3$. In these figures, the red triangles are obtained from MC-Glauber distribution and the error bars indicate the statistical errors. Here, the centrality parameter between $0$ to $80\%$ is divided into 16 bins and 14\,000 events are generated in each bin by iEBE-VISHNU  (see Sec.~\ref{simulation} for more details).
In the same figures mentioned above, the blue dashed curve demonstrates the cumulants extracted from the elliptic-power distribution. To find them, we have used Eqs.~\eqref{b4}$-$\eqref{b6} with $\alpha$ and $\ve_0$ obtaining from the initial states fit. 

 Note that by fixing two inputs from MC-Glauber, namely $\alpha$ and $\ve_0$,  we have found a large number of outputs which is a set of cumulants. In Fig.~\ref{n2cumulants} and Fig.~\ref{n3cumulants}, one can see a qualitative agreement between standardized cumulants
obtained from MC-Glauber comparing to that acquired from the elliptic-power distribution.  

In addition,  we have $\hat{\E}^{(2)}_{01}\simeq\hat{\E}^{(2)}_{11}\simeq\hat{\E}^{(2)}_{01}\simeq\hat{\E}^{(2)}_{21}\simeq\hat{\E}^{(2)}_{03}\simeq\hat{\E}^{(2)}_{31}\simeq\hat{\E}^{(2)}_{13}\simeq0$ for $n=2$, in agreement with the symmetries of the elliptic-power distribution. Moreover, we checked that for MC-Glauber model $\hat{\E}^{(2)}_{12}\sim\hat{\E}^{(2)}_{22}\sim 0$.

For $n=3$, all the cumulants extracted from elliptic-power distribution up to order three are equal to zero. This feature is also observed for the cumulants extracted from MC-Glauber. We have checked that it is a reasonable assumption to consider $\hat{\E}^{(3)}_{22}\sim 0$. Also we can see from the Fig.~\ref{n3cumulants} that the cumulants $\hat{\E}^{(3)}_{40}$ and $\hat{\E}^{(3)}_{04}$ are almost equal. Note that it is an exact equality for the cumulants extracted from elliptic-power distribution. In other words, except $\hat{\E}_{20}^{(3)}$ and $\hat{\E}_{02}^{(3)}$, there is only one independent standardized cumulant in $n=3$ harmonics, $\hat{\E}^{(3)}_{40} \sim \hat{\E}^{(3)}_{04},$
in agreement with the elliptic-power distribution.

\subsection{Cumulants of collision final state}

The momentum distribution of the particles observed in the detector is correlated with the heavy-ion collision initial state. Here, we will try to clarify the relation between the cumulants obtained from the initial distribution and the final-state particle distribution. 

The azimuthal distribution of particles is analyzed via Fourier series,
\begin{eqsplit}\label{FourierAnal}
	\frac{2\pi}{N}\frac{dN}{d\phi}= 1+\sum_{n=1}^{\infty}2v_n\cos\left[n(\phi-\psi_n)\right].
\end{eqsplit}
Defining the complex flow harmonics $\boldsymbol{v}_n=v_n e^{in\psi_n}$, we can find $\boldsymbol{v}_n=\la e^{in\phi} \ra_s$, where $\la\cdots\ra_s$ is averaging in a single event. Instead of using the complex form of the flow harmonics, we occasionally use the Cartesian form of them defined as follows:
\bea\label{polarCoord}
v_{n,x}=v_n\cos(n\psi_n),\quad v_{n,x}=v_n\sin(n\psi_n).
\eea
Each nonvanishing $\boldsymbol{v}_n$ measures how nonuniform the final particle distribution is. For example, the ellipticity of the initial state produced in the noncentral collisions is manifested in the nonzero values for $v_{2,x}$ and $v_{2,y}$.

In the experiment,  the azimuthal angle of the reaction plane $\phi_{\text{RP}}$ is not a direct observable which means the symmetry angle $\psi_n$ is unknown.  Although the angle $\psi_n$ is not known,  we are still able to find the parameter $v_n$ in each centrality  class by studying the $2q$-particle correlation functions $c_n\{2q\}$ \cite{Borghini:2000sa,Borghini:2001vi}.\footnote{ These correlation functions are written in terms of $\la\la e^{i\,n(\phi_1+\cdots+\phi_q-\phi_1-\cdots-\phi_q)}\ra_s\ra$, where $\phi_i$ is the azimuthal angle of a particle in a given event. First, the average is performed over a single event and after that we average the results over many events \cite{Borghini:2000sa,Borghini:2001vi}.  }
In Ref.~\cite{Borghini:2001vi}, the relation between $2q$-particle correlation functions and $v_n$ is found,
\bea\label{nParticleCorrelation}
\sum_{q} \frac{\lambda^{2q}}{(q!)^2}c_n\{2q\}=\log I_0(\lambda v_n).
\eea
In the following, we refer to $v_n$ obtained by equating the coefficients of $\lambda^{2q}$ in two sides as $v_n\{2q\}$\footnote{From \eqref{nParticleCorrelation}, one finds explicitly
	\bea
	v_n^2\{2\}\equiv c_n\{2\},\quad v_n^4\{4\}\equiv -c_n\{4\},\quad v_n^6\{6\}\equiv c_n\{6\}/4
	\eea}. 

 The complexity of the initial energy density and its fluctuation from one event to the other leads to different values for $v_{n,x}$ and $v_{n,y}$, even for the events in the same centrality class. One should note that the reaction plane angle is under control in simulations and we can set it to zero ($\phi_{\text{RP}}=0$). As a result,  we find a distribution $p(v_{n,x},v_{n,y})$ for an ensemble of events due to the event-by-event fluctuations in simulation. In this case, one can find the cumulants of this distribution via \eqref{cumulGen} similar to what we have done for $\ve_n$ distribution. We will refer to the cumulants extracted from  $p(v_{n,x},v_{n,y})$  as $\V_{kl}^{(n)}$.

In order to relate the  $c_n\{2q\}$ with the cumulant $\V_{kl}^{(n)}$, one has to integrate over $\psi_n$ in \eqref{cumulGen} first and then compare it with \eqref{nParticleCorrelation} \cite{Borghini:2001vi,Giacalone:2016eyu}. \blk We can set $\lambda_x=\lambda \cos \psi_n$ and $\lambda_y=\lambda \sin \psi_n$ in \eqref{cumulGen} and define the generating function $G(\lambda)$ as
\bea
\log G(\lambda)=\log \left(\int_{0}^{2\pi}\frac{d\psi_n}{2\pi}\la e^{\lambda(v_{n,x}\cos\psi_n+v_{n,y}\sin\psi_n)} \ra\right).\nn\\
\eea
Consequently, one can find the relation between $c_n\{2q\}$ (or $v_n\{2q\}$) with $\V_{pq}^{(n)}$ by equating the expansions of $\log I_0(\lambda v_n)$ and $\log G(\lambda)$. 

After some calculations, one simply finds that the general form of the $c_n\{2k\}$ in terms of $\V_{pq}$ has the following structure\footnote{We occasionally ignore the superscript $(n)$ in the 2D cumulants for simplicity in notation. } \footnote{In the following, we present two explicit examples:
	\bea
	c_n\{2\}&=&\V_{01}^2+\V_{10}^2+\V_{02}+\V_{20},\nn\\
	c_n\{4\}&=&-\V_{01}^4-2 \V_{10}^2 \V_{01}^2+2 \V_{02} \V_{01}^2-2 \V_{20} \V_{01}^2+4 \V_{03} \V_{01}\nn\\
	&+&8 \V_{10} \V_{11} \V_{01}+4 \V_{21} \V_{01}-\V_{10}^4+\V_{02}^2-2 \V_{02} \V_{10}^2+4 \V_{11}^2  \nn\\
	&+&\V_{20}^2+\V_{04}+4 \V_{10} \V_{12}+2 \V_{10}^2 \V_{20}-2 \V_{02} \V_{20}+2 \V_{22}\nn\\
	&+&4 \V_{10} \V_{30}+\V_{40}.\nn
	\eea}:
\bea\label{explicitCn}
c_n\{2k\}=\sum_{\{\ell_i,p_i,q_i\}} a_{\{\ell_i,p_i,q_i\}}\V_{p_1q_1}^{\ell_1}\cdots \V_{p_Nq_N}^{\ell_N},
\eea
where
\bea \label{constraint}
\sum_{i=1}^{N}\ell_i(p_i+q_i)=2k,
\eea
and $a_{\{\ell_i,p_i,q_i\}}$ is a real number.

The constraint \eqref{constraint} is found by the following argument. Clearly, $\V_{mn}$,  which is  given by \eqref{cumulGen}, is homogeneous. Consider the rescaling $\xi_i\to \chi \xi_i$ for $i=x,y$, where $\chi$ is a real number. Then one can assume that $\xi_i$ is unchanged while $\lambda_i$ is replaced by $\chi \lambda_i$ in the left-hand side of Eq.~\eqref{cumulGen}. Now in order to find same cumulant by equating two sides of the equation, we need to have $\A_{mn}\to\chi^{m+n} \A_{mn}$. Recall that $c_n\{2k\}$ can be obtained by integrating the azimuthal angle of the left-hand side of \eqref{cumulGen}. As a result, we expect similar scaling for them, $c_n\{2k\}\to \chi^{2k}c_n\{2k\}$. By using these ingredients, one can find the constraint \eqref{constraint}. The same argument will be used to  find the relation between cumulants obtained from $\boldsymbol{\ve}_n$ and $\boldsymbol{v}_n$ distributions, considering the hydrodynamic linear response [see \eqref{linear} and \eqref{linearCumul}].   

 Due to the averaging over $\psi_n$, there is more information about event-by-event fluctuations in $\V_{pq}^{(n)}$ compared to the $c_n\{2k\}$. If one obtains $c_n\{2k\}$ explicitly in terms of $\V_{pq}^{(n)}$, then it can be seen that the number of terms in $c_n\{2k\}$  grows rapidly with increasing $k$. In the following section, we would like to find the informations encoded in   $c_n\{2k\}$ from $\V_{pq}^{(n)}$ as much as possible. Then we will argue how to truncate $c_n\{2k\}$ expansion.

\section{2D Standardized Cumulants from Correlation Functions}\label{toyModelSection}

In the previous section, we showed that, in principle, one would be able to obtain $c_n\{2k\}$ in terms of cumulants $\V_{pq}^{(n)}$. The former is an experimental observable while the latter can be obtained from simulation. On the other hand, one knows that the elliptic-power distribution can explain the distribution of $\boldsymbol{\ve}_n$ obtained from more sophisticated initial condition models. This distribution leads to a semianalytical result for $\hat{\E}_{pq}^{(n)}$. The semianalytical $\hat{\E}_{pq}^{(n)}$ can be considered as $\hat{\V}_{pq}^{(n)}$ by using the hydrodynamic linear response approximation. Consequently, we may use the elliptic-power distribution as a toy model to find a reasonable approximation for truncating the expansion \eqref{explicitCn}.

The hydrodynamic response to the initial state has been studied from different directions \cite{Teaney:2010vd,Gardim:2011xv,Teaney:2012ke,Gardim:2014tya,Yan:2015jma,Noronha-Hostler:2015dbi}.  However, it is a reasonable approximation for $n=2,3$ to consider the hydrodynamic response being linear \cite{Teaney:2010vd,Gardim:2011xv,Gardim:2014tya,Noronha-Hostler:2015dbi},
\bea\label{linear}
\boldsymbol{v}_n \simeq \chi_n \boldsymbol{\ve}_n,
\eea
where $\chi_n$ is a real valued constant of proportionality. With this approximation and using the homogeneity of cumulants, we immediately find
\bea\label{linearCumul}
\V_{pq}^{(n)} \simeq \chi_n^{p+q} \E_{pq}^{(n)}.
\eea
Referring to the definition of the standardized cumulants \eqref{normalCumul}, we see that at the linear approximation 
\bea\label{linearApprox}
\hat{\V}_{pq}^{(n)}\simeq \hat{\E}_{pq}^{(n)}.
\eea

Let us note that in \eqref{explicitCn}, the $2q$-particle correlation function, $c_n\{2q\}$, was given in terms of $\V_{kl}^{(n)}$. In this subsection, we exploit the  equation  \eqref{linearApprox} and rewrite $c_n\{2q\}$ in terms of $\hat{\E}_{kl}^{(n)}$. 

\begin{figure}[t!]
	\begin{center}
		\begin{tabular}{c}
			\includegraphics[scale=0.4]{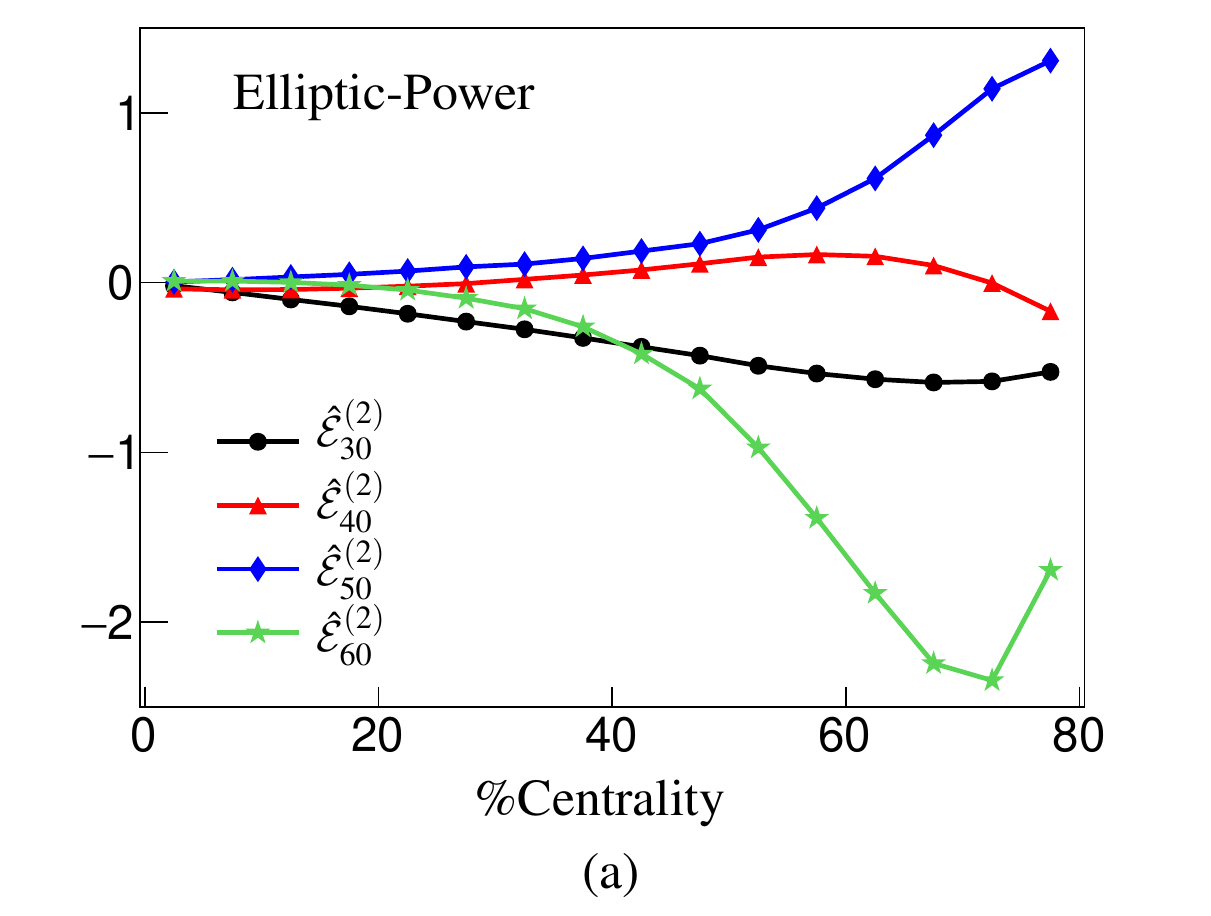}			\vspace*{1.0cm}\\
			\includegraphics[scale=0.4]{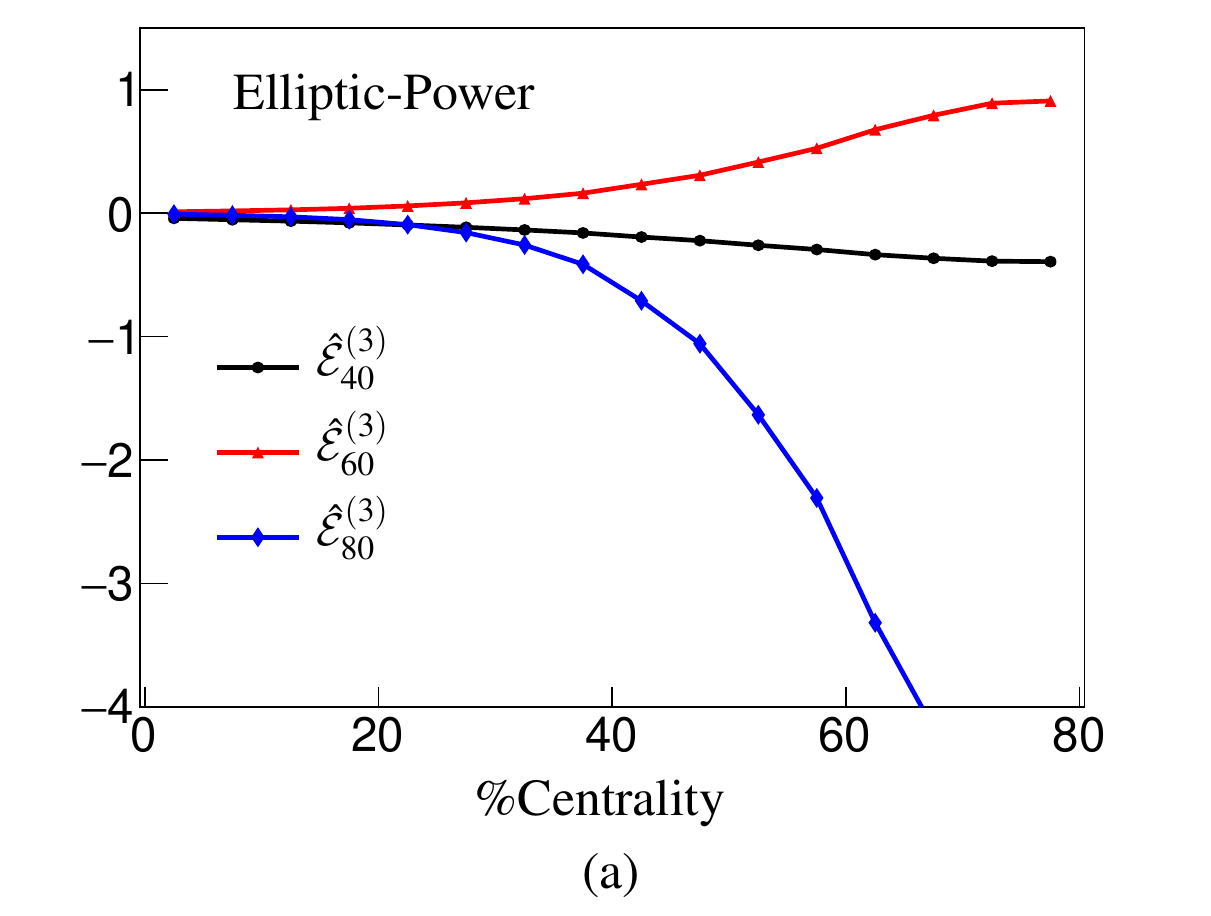}					
		\end{tabular}
		\caption{(Color online) A few leading standardized cumulants obtained from elliptic-power distribution for (a) $n=2$ and (b) $n=3$. Typically by increasing the order, the value of the cumulants increase. }
		\label{cumulantCompare}
	\end{center}
\end{figure}

In order to find general form of $c_n\{2q\}$ in terms of $\hat{\E}_{kl}^{(n)}$, the following remarks must be considered:
\begin{itemize}
	\item In elliptic-power distribution, one can check that for $n=2$ it is a good approximation to consider $\E_2^{(n)}\equiv\E_{02}^{(n)}\simeq \E_{20}^{(n)}$. It turns out that for $n=3$, this relation becomes exact.
	\item An explicit calculation shows that terms $\hat{\E}_{10}^{2k-2}\hat{\E}_{02}$ has the same coefficient as $\hat{\E}_{10}^{2k-2}\hat{\E}_{20}$ but with opposite sign\footnote{We checked it for $k=2,3,4,5,6$.}. Also for the elliptic-power distributions, the cumulant $\hat{\E}^{(n)}_{11}=0$.
\end{itemize}
Using all the above considerations together with \eqref{explicitCn}  [and \eqref{normalCumul}], one finds
\begin{eqsplit}\label{cnExpansion}
	\frac{c_n\{2k\}}{\chi_n^{2k}\E_2^k}&\simeq \#_1\hat{\E}_{10}^{2k}+\hat{\E}_{10}^{2k-3}(\#_2 \hat{\E}_{30}+\#_3 \hat{\E}_{12}+\cdots)\\&+ \#_4\hat{\E}_{10}^{2k-q}\hat{\E}_{q,0}+\cdots+\#_l\hat{\E}_{2k-q',q'}+\cdots.
\end{eqsplit}
Recall that due to the symmetries of elliptic-power distribution, all the standardized cumulants $\hat{\E}^{(2)}_{k,2q+1}$ are zero for $n=2$ while  for $n=3$, the only nonzero cumulants are $\hat{\E}^{(3)}_{2k,2q}$.

In Fig.~\ref{cumulantCompare}, we have compared a number of cumulants extracted from elliptic-power distribution. Obviously, the nonvanishing $\hat{\E}_{pq}$ for $p+q\geq 3$ indicates that the elliptic-power distribution is not Gaussian. So we may describe the elliptic-power distribution by the two dimensional Gram-Charlier A series. Let us briefly explain how it works.

A general two dimensional distribution $\PP(\xi_x,\xi_y)$ can be written as (Appendix \ref{RGCD}) 
\bea\label{gramCharlier}
\PP(\xi_x,\xi_y)\simeq\frac{1+\mathcal{H}}{2\pi \sqrt{\A_{20}\A_{02}}} e^{-\frac{(\xi_x-\A_{10})^2}{2\A_{20}}-\frac{(\xi_y-\A_{01})^2}{2\A_{02}}}
\eea
where 
\bea\label{expansion}
\mathcal{H}=\sum_{\substack{m=n=1,\\ m+n\geq 3}}   \frac{h_{mn}}{m!n!} He_n(\frac{\xi_x-\A_{10}}{\sqrt{\A_{20}}})He_m(\frac{\xi_y-\A_{01}}{\sqrt{\A_{02}}}).\quad
\eea
In the above, $He_n$ is the (probabilistic) Hermite polynomial  and $h_{mn}=\hat{\A}_{mn}$ for $m+n\leq 5$. Let us emphasize that for rotationally symmetric distributions (e.g., odd flow harmonic distributions), we have $\hat{\A}_{mn}=0$ for odd $m+n$. Referring to \eqref{hmnSix}, we deduce that for such case, $h_{mn}=\hat{\A}_{mn}$ even if $m+n=6$.  The Eq.~\eqref{gramCharlier} is the two-dimensional Gram-Charlier A series. For Gaussian distributions, $\mathcal{H}=0$ while for each non-Gaussian distribution, a certain set of coefficients $\hat{\A}_{mn}$ have nonzero values, therefore, $\mathcal{H}\neq 0$. In the following, we study $\hat{\A}_{mn}$ associated with elliptic-power distribution by replacing $\hat{\A}_{mn}$ with $\hat{\E}_{mn}$ in \eqref{gramCharlier}.

As can be seen in Fig.~\ref{cumulantCompare}, typically by increasing $p+q$ the value of $\hat{\E}_{pq}$ increases. It has been checked that the increase rate of $\hat{\E}_{pq}$ is smaller than that  of $p!q!$.  As a result, the coefficients of successive terms in the expansion \eqref{expansion} are decreasing. This means that they are less important in the non-Gaussian shape of the distribution.

\begin{figure*}[t!]
	\begin{center}
		\begin{tabular}{cccc}
			\includegraphics[scale=0.28]{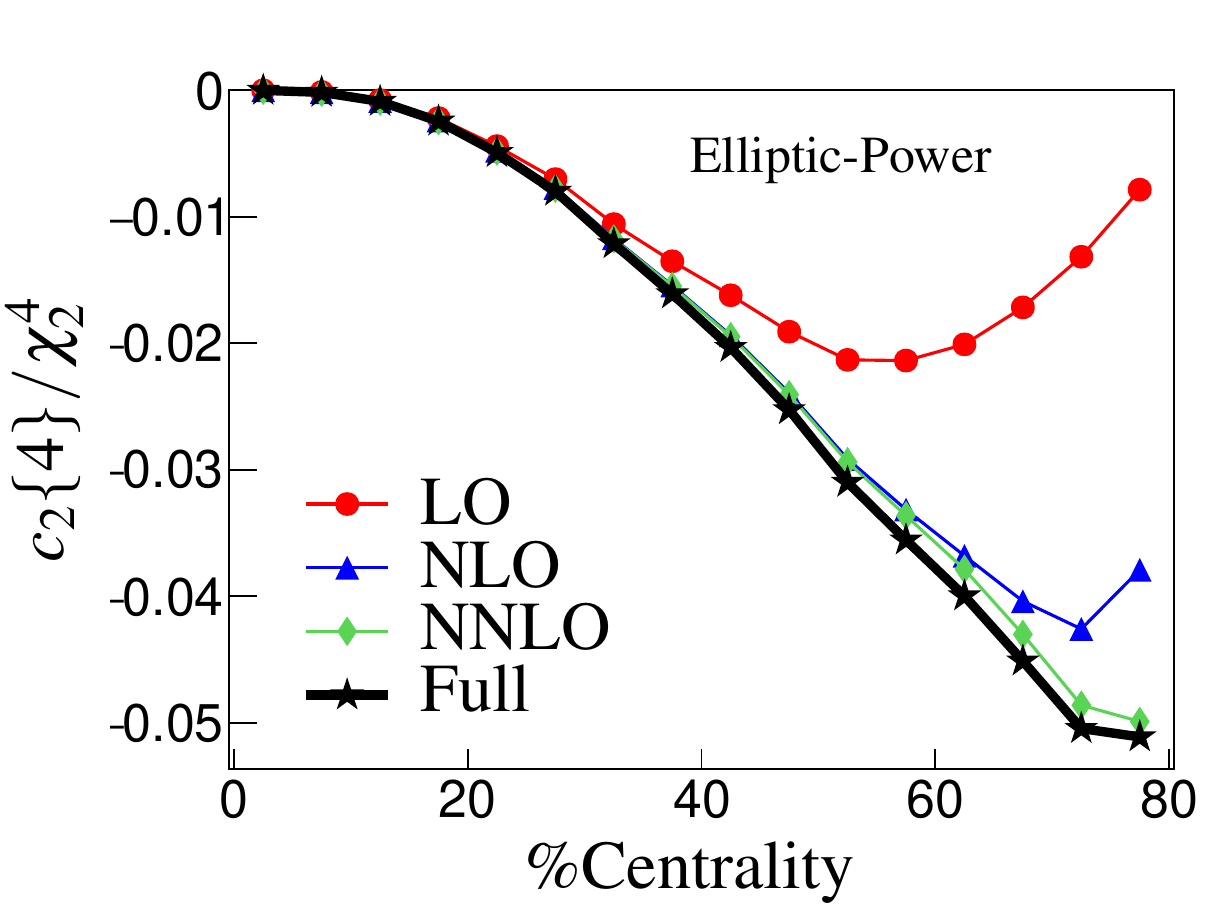}\hspace*{0.2cm}		
			& 			\includegraphics[scale=0.28]{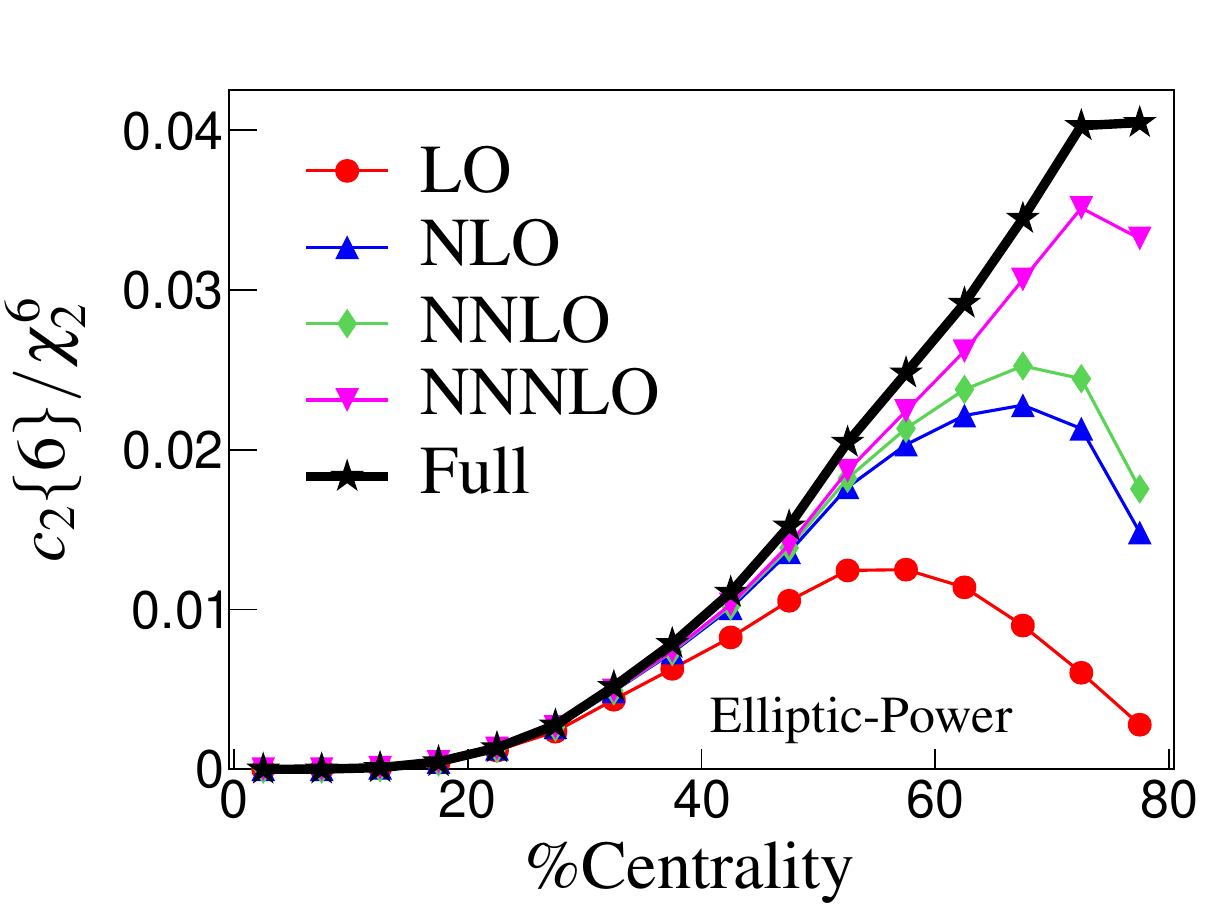}\hspace*{0.2cm}		
			& 			\includegraphics[scale=0.28]{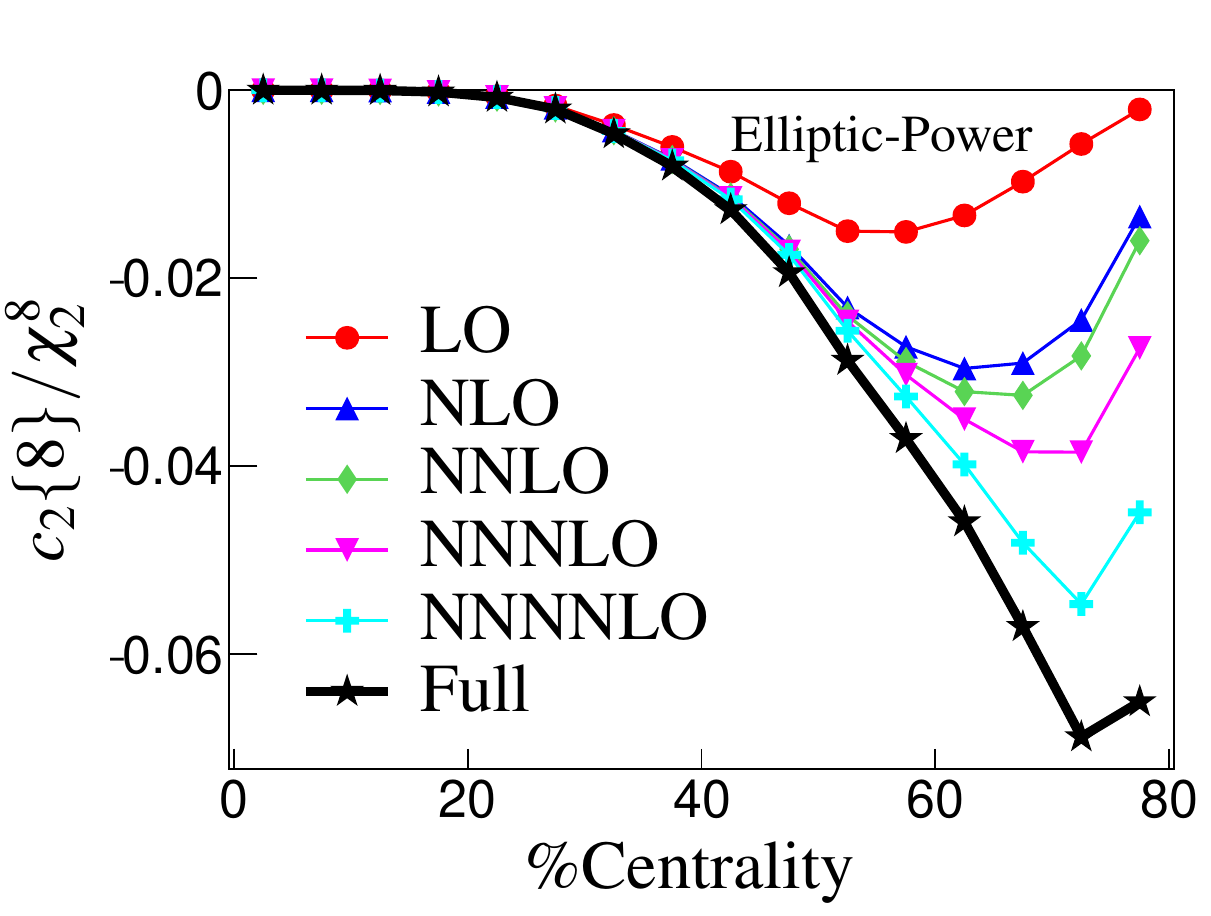}		
		\end{tabular}		
	\end{center}
	\caption{(Color online) Comparing different values of $c_2\{4\}$,  $c_2\{6\}$, and $c_2\{8\}$ in the standardized cumulant expansion truncation.  }
	\label{truncation}
\end{figure*}

 Comparing the order of magnitude of different $\hat{\E}_{k l}^{(n)}$'s to each other, there is an exception for the second harmonics.  In $n=2$, the nonzero value for $\hat{\E}_{10}$ comes from the ellipticity of the initial condition in noncentral collisions  and its   value is relatively larger than the cumulants originated  from the event-by-event fluctuations (see Fig.~\ref{n2cumulants}). In this case, we expect the terms to be  ordered with decreasing power of $\hat{\E}_{10}$. The leading order (LO) comes from $\hat{\E}_{10}^{2k}$ and the next to leading order (NLO) is\footnote{As we explained earlier, $\hat{\E}_{pq}$ increases by increasing $p+q$.  As a result (for $n=2$), it is probable that if we study higher-order cumulants, then the terms contain higher-order cumulants (small power of $\hat{\E}^{(2)}_{10}$) becomes dominant. We checked this point for the distribution under consideration and we found it is not the case for $c_2\{2\}$, $c_2\{4\}$, $c_2\{6\}$, and $c_2\{8\}$. In other words, expansion with decreasing power $\hat{\E}^{(2)}_{10}$ is reliable for $c_2\{2k\}$, $k\leq 4$. } 
$$\hat{\E}_{10}^{2k-3}(\#_1 \hat{\E}_{30}+\#_2 \hat{\E}_{12}+\cdots).$$ 

As we mentioned earlier, the distribution of $\ve_2$ is not rotationally symmetric. We have also seen that it is skewed in the $\ve_{2,x}$ direction [see Fig.~\ref{EllipticPowerFit}(a)]. In what follows, we argue that the other higher-order cumulants of $\ve_2$ distribution cannot be extracted from $c_2\{2k\}$ truncation.

Let us emphasize that we only keep terms with decreasing power of $\hat{\E}^{(2)}_{10}$ in $c_2\{2k\}$ expansion. In order to get fairly accurate result up to $60\%$ centrality for $c_2\{2k\}$, $k=2,3,4$ (see Fig.~\ref{truncation}), we have to use the following truncations:
\bea
\frac{c_2\{2\}}{\chi_2^2 \E_2}&\simeq&\hat{\E}_{10}^2+2\label{c22Elliptic}\\
\frac{c_2\{4\}}{\chi_2^4 \E_2^2}&\simeq&-\hat{\E}_{10}^4+4\hat{\E}_{10}\left(\hat{\E}_{30}+\hat{\E}_{12}\right)\nn\\
&+& \hat{\E}_{40}+\hat{\E}_{22}+\hat{\E}_{04}\label{c24Elliptic}\\
\frac{c_2\{6\}}{\chi_2^6 \E_2^3}&\simeq&4\hat{\E}_{10}^6-8\hat{\E}_{10}^3\left(2\hat{\E}_{30}+3\hat{\E}_{12}\right)\nn\\
&+&6 \hat{\E}_{10}^2 (\hat{\E}_{40}-\hat{\E}_{04})\label{c26Elliptic}\\
&+&6\hat{\E}_{10}\left(\hat{\E}_{50}+2\hat{\E}_{32}+\hat{\E}_{14}\right)\nn\\
\frac{c_2\{8\}}{\chi_2^8 \E_2^4}&\simeq& -33 \hat{\E}_{10}^8+24\hat{\E}_{10}^5\left(7\hat{\E}_{30}+11\hat{\E}_{12}\right)\nn\\
&-&\hat{\E}_{10}^4(62\hat{\E}_{40}+12\hat{\E}_{22}-66\hat{\E}_{04})\nn\\
&-&8\hat{\E}_{10}^3(5\hat{\E}_{50}+14\hat{\E}_{32}-9\hat{\E}_{14})\label{c28Elliptic}\\
&-&12\hat{\E}_{10}^2(\hat{\E}_{60}+\hat{\E}_{42}-14\hat{\E}_{30}^2\nn\\
&&\hspace*{1.8cm}-44\hat{\E}_{30}\hat{\E}_{12}-\hat{\E}_{24}-30\hat{\E}_{12}^2-\hat{\E}_{06}).\nn
\eea

Figure \ref{truncation} displays the exact and approximate values for $c_2\{4\}$, $c_2\{6\}$, and $c_2\{8\}$ obtained from the elliptic-power distribution. Here, we did not plot $c_2\{2\}$ because the relation \eqref{c22Elliptic} is almost exact with the only approximation $\E_{20}\simeq\E_{02}$. In the same figure, by moving from $c_2\{4\}$ to $c_2\{8\}$ more terms are needed to find a good approximation compared to the exact relation.

These observations are in agreement with the results of Ref.\cite{Giacalone:2016eyu}. In Ref.~\cite{Giacalone:2016eyu}, only the NLO terms, i.e., the contributions in the first line in each of equations \eqref{c22Elliptic} to \eqref{c28Elliptic}, have been considered. By use of this approximation, the authors of Ref.~\cite{Giacalone:2016eyu}, computed $\E_{30}^{(2)}$, considering $c_2\{2\}$, $c_2\{4\}$, and $c_2\{6\}$.\footnote{The approximate $c_2\{2\}$, $c_2\{4\}$, and $c_2\{6\}$ used in Ref.~\cite{Giacalone:2016eyu}, have been depicted by blue curves in Fig.~\ref{truncation}.} Their results  are in agreement with experimental data. It is worth mentioning that the approximation they used is obtained by studying a full hydrodynamic simulation.

Note that if we are interested in finding cumulants beyond skewness, $c_2\{8\}$ is needed to be taken into account. However, as can be seen from Fig.~\ref{truncation}, going from NLO to NNLO does not improve the accuracy of $c_2\{6\}$ and $c_2\{8\}$ remarkably. In other words, it would not be easy to find the standardized cumulants beyond the skewness for the elliptic flow distribution.

For $n=3$, all the nonzero cumulants are coming from the event-by-event fluctuations and $\hat{\E}_{10}$ is zero due to the symmetry. As we observed in Fig.~\ref{cumulantCompare}(b), we expect the leading term of $c_3\{2k\}$ to be $\hat{\E}_{pq}$ with $p+q=2k$. Let us note that although $\hat{\E}_{pq}$ with $p+q=2k$ has the main contribution to $c_3\{2k\}$, it does not seriously affect the deviation of the distribution from Gaussianity.

 Additionally,  unlike the $\ve_2$ distribution case, for $\ve_3$ the distribution is rotationally symmetric in the $(\ve_{3,x}$-$\ve_{3,y})$ plane (see Fig.~\ref{EllipticPowerFit}(b) and Eq.\eqref{powerDist}). As a result, $\ve_3$ distribution is not skewed, however, it can have a nonzero kurtosis in the radial direction.  In $n=3$, we calculate a number of nonzero cumulants, including kurtosis, in the radial direction.

Considering \eqref{cnExpansion} for $c_3\{2k\}$ expansion and the previously mentioned properties of $\E_{pq}^{(3)}$ for elliptic-power distribution, one finds
\bea
\K_2&\equiv& \frac{c_3\{2\}}{\chi_2^2 }= \E_{20}+\E_{02},\label{k2}\\
\K_4&\equiv&\frac{c_3\{4\}}{\chi_2^4 }=\E_{40}+2\E_{22}+\E_{04}, \label{k4}\\
\K_6&\equiv&\frac{c_3\{6\}}{\chi_2^6 }=\E_{60}+3\E_{42}+3\E_{24}+\E_{06}.\label{k6}
\eea 
Note that the relations \eqref{k2}$-$\eqref{k6} are exact, by this we mean that we have not used any truncation when deriving them. However, for distributions obtained from more realistic models (e.g., MC-Glauber), the above relations are truncations of expansion \eqref{cnExpansion} and so approximately true.

 In order to show the relation between the cumulants in the radial direction and $\K_q$, let us use the polar coordinate $\ve_{3,x}=\ve_3\cos\varphi$ and $\ve_{3,y}=\ve_3\sin\varphi$. Doing so, we obtain\footnote{In order to clearly distinguish between averaging over elliptic-power and power distributions, we use the subscripts EP and P, respectively.  },
 \bea\label{2Dto1Dradial}
 \la \ve_{3,x}^m\ve_{3,y}^n \ra_{EP}=\la \ve_3^{m+n}\ra_{P}\int \frac{d\varphi}{2\pi} \cos^m\varphi \sin^n\varphi.
 \eea
In this equation, the average in the left-hand side has been taken by the distribution function \eqref{EPD} while for the average in the right-hand side, the distribution \eqref{powerDist} has been used. In general, for any rotationally symmetric distribution, the averaging in the azimuthal integration is factorized and the moments with either odd $m$ or odd $n$ vanish.

In the right-hand side of the equations \eqref{k2}$-$\eqref{k6}, the cumulants $\E_{pq}^{(3)}$ have been written in terms of moments $\la \ve_{3,x}^m\ve_{3,y}^n \ra_{EP}$.  One can substitute \eqref{2Dto1Dradial} into \eqref{k2}$-$\eqref{k6} to find $\K_n$ in terms of moment $\la \ve_3^q\ra_P$. As an example,
\bea
\K_2&=& \la \ve_3^2 \ra_P,\nn\\
\K_4&=&\la \ve_3^4\ra_P  -3\la \ve_3^2 \ra_P^2.\nn
\eea

On the other hand, the cumulants $\kappa_2$ and $\kappa_4$ (introduced in Sec.~\ref{terminologu}) of the one dimensional power distribution \eqref{powerDist} is given by
\bea
\kappa_2&=& \la \ve_3^2 \ra_P-\la \ve_3 \ra_P^2,\nn\\
\kappa_4&=&\la \ve_3^4\ra_P -4\la \ve_3^3\ra_P \la \ve_3\ra_P -3\la \ve_3^2 \ra_P^2 +12 \la \ve_3^2 \ra_P \la \ve_3 \ra_P^2\nn\\
& -&6\la \ve_3 \ra_P^4.\nn
\eea
 In fact, $\K_n$ coincides with $\kappa_n$ if the moments $\la \ve_3^{2q+1}\ra$ are removed. This actually happens for every rotationally symmetric distribution due to the $\varphi$ integral in \eqref{2Dto1Dradial}. As a result, the standardized cumulants of such distribution may be written in terms of $\K_q$ as follows:
\bea\label{bigGamma}
\Gamma_{q-2}=\frac{\K_{q}}{\K_2^{q/2}}.
\eea
For instance, $\Gamma_2$ is the kurtosis. In this case, the skewness, $\Gamma_1$, is zero because $\K_3=0$.

Rotational symmetry suggests to integrate over the azimuthal angle in \eqref{gramCharlier}. To do so, we change the variable $(\xi_x,\xi_y)$ to $(\xi_r,\xi_{\phi})$ with $\xi_r=(\xi_x^2+\xi_y^2)^{1/2}$ and $\xi_{\phi}=\text{atan2}(\xi_y/\xi_x)$. Using \eqref{linear} and after some cumbersome calculations, one obtains (see Appendix \ref{RGCD} for more details)
\begin{eqsplit}\label{aDistribution}
	p(v_3)=&\left[1+\Gamma_2\Q_4(\frac{v_3}{v_3\{2\}})+\Gamma_4\Q_6(\frac{v_3}{v_3\{2\}})+\cdots\right]\\
	&\hspace*{2.5cm}\times\frac{2 v_3}{v_3^2\{2\}}\exp\left[-\frac{v^2_3}{ v_3^2\{2\}}\right]
\end{eqsplit}
where 
\bea
\Gamma_2&=&(\hat{\V}_{40}+2\hat{\V}_{22}+\hat{\V}_{04})/4,\label{36}\\
\Gamma_4&=&(\hat{\V}_{60}+3\hat{\V}_{42}+3\hat{\V}_{24}+\hat{\V}_{06})/8,\label{37}
\eea
and
\bea
\Q_4(\xi)&=&\frac{1}{4}\left[\xi^4-4\, \xi^2+2\right],\label{Q4}\\
\Q_6(\xi)&=&\frac{1}{36}\left[\xi^6-9\,\xi^4+18\,\xi^2-6\right]\label{Q6}.
\eea
By using Eq.~\eqref{linearCumul} together with Eqs.~\eqref{k2}$-$\eqref{k6} we find
\bea
\Gamma_2&=&-\left(\frac{v_3\{4\}}{v_3\{2\}}\right)^4,\label{expKurt}\\
\Gamma_4&=&4\left(\frac{v_3\{6\}}{v_3\{2\}}\right)^6.\label{expSixOrder}
\eea
We call the distribution \eqref{aDistribution} \textit{Radial-Gram-Charlier} (RGC) distribution. Here, the random variable is $v_3$ while $v_3\{2\}$, $\Gamma_2$, and $\Gamma_4$ are constants that can be obtained by a fitting process.\footnote{We are able to obtain these quantities by computing $c_n\{2k\}$, too.} Note that if we set $\Gamma_2=\Gamma_4=0$, then the Gaussian distribution is found.  We would like to note that the ratio $v_n\{4\}/v_n\{2\} $ has been used recently to study  the fluctuations of different initial condition models based on the hydrodynamic linear response approximation  \cite{Giacalone:2017uqx}. For $n=3$, this ratio is equal to $(-\Gamma_2)^{1/4}$.

In this section, we studied the reasonable truncation of $2q$-particle correlation cumulant expansion by exploiting a semianalytical model. More importantly, we found a new parametrization for the distribution $p(v_3)$  which describes the leading deviation of $v_3$ distribution from Gaussian distribution, with two parameters, namely $\Gamma_2$ and $\Gamma_4$. The results of model we used in this section (elliptic-power together linear hydrodynamic response) are not too reliable to be compared with the experimental data. For this reason, in the next section, we use a more realistic model, i.e., the iEBE-VISHNU event generator together with MC-Glauber model. To compare with experimental data, we then apply the truncations obtained in the current section to the mentioned model, using also the RGC distribution.

\section{MC-Glauber Model and Beyond Hydrodynamic Linear Response}\label{simulation}

\begin{figure*}[th]
	\begin{center}
		\begin{tabular}{cccc}
			\includegraphics[scale=0.28]{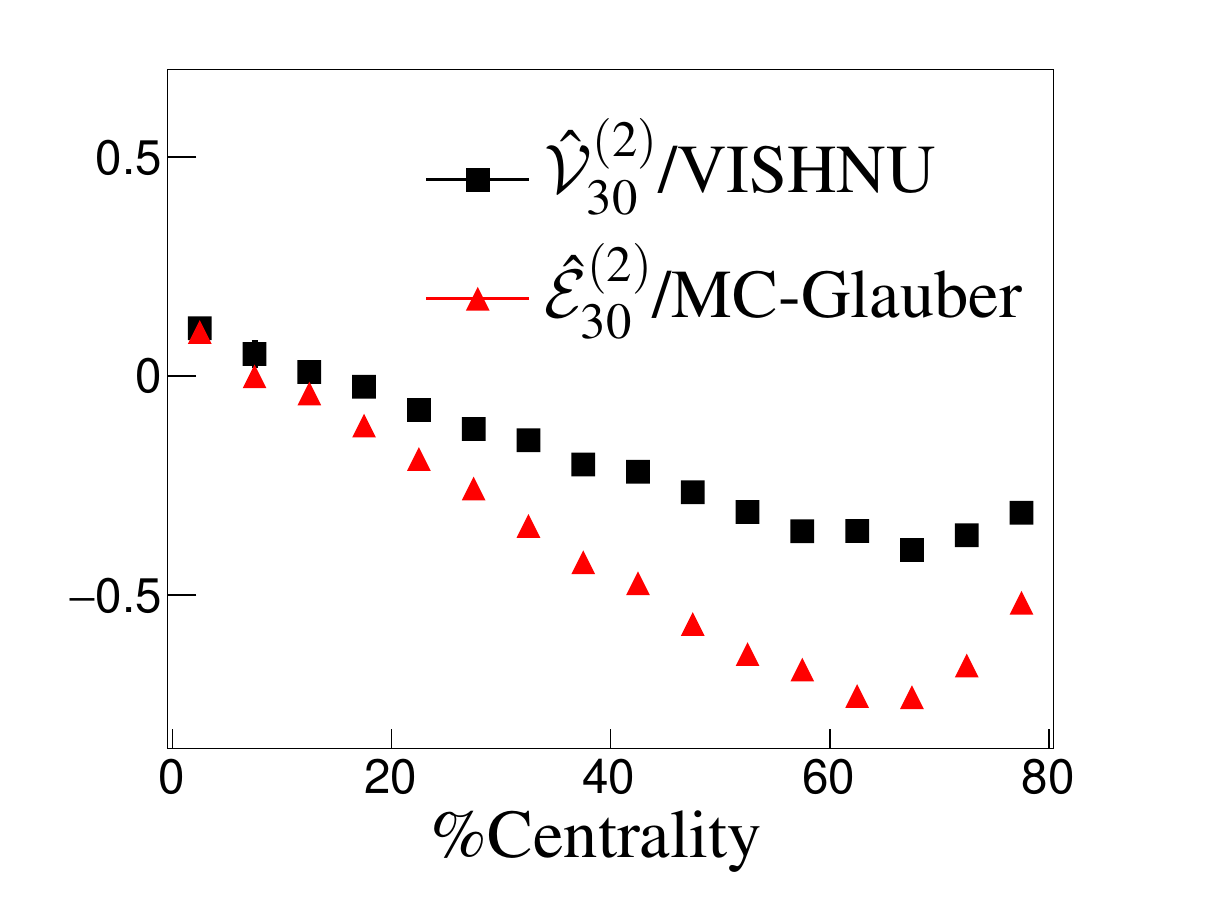}\hspace*{0.5cm}		
			& 			\includegraphics[scale=0.28]{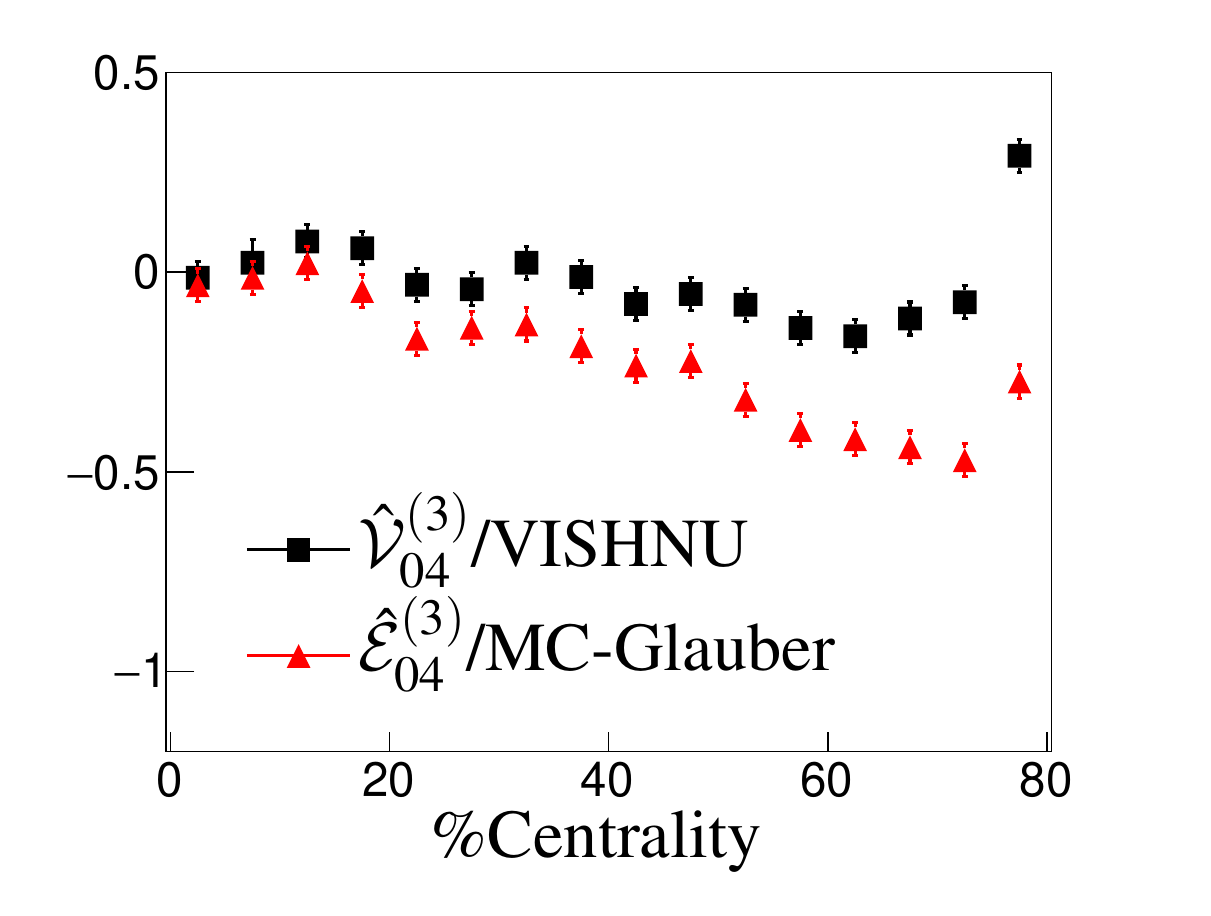}\hspace*{0.5cm}		
			& 			\includegraphics[scale=0.28]{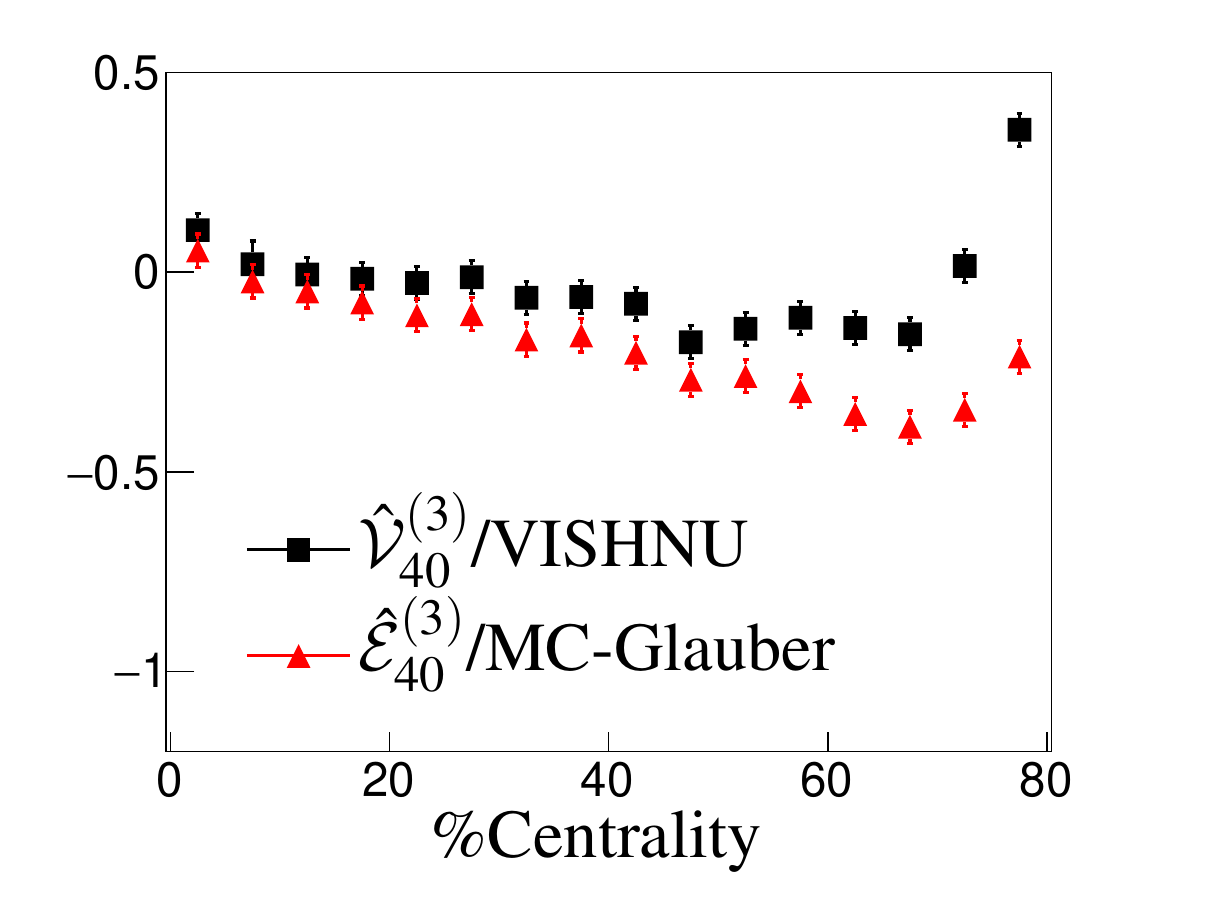}		
		\end{tabular}									
		\normalsize 		
	\end{center}
	\caption{(Color online) Some nonzero standardized cumulants of $p(\ve_{n,x},\ve_{n,x})$ and $p(v_{n,x},v_{n,x})$. }
	\label{NonLinearResponse}
\end{figure*}

The skewness of $v_2$ distribution has been calculated in Ref.~\cite{Giacalone:2016eyu} by using the viscous relativistic hydrodynamical  code V-USPHYDRO \cite{Noronha-Hostler:2013gga,Noronha-Hostler:2014dqa,Noronha-Hostler:2015coa}. While in the same reference, the skewness has been also found from experimental data, nothing has been mentioned about $v_3$ distribution there. In the current section, we focus on finding the standardized cumulants of $v_3$ distribution.

Here, we use the heavy-ion collision event generator iEBE-VISHNU \cite{Shen:2014vra} to study the evolution of the initial state generated by the MC-Glauber model (implemented in iEBE-VISHNU). After generating the initial condition, we let it evolve through a 2+1 dimensional viscous hydrodynamic model based on the causal Israel-Stewart formalism. At the end of the hydrodynamic evolution, each fluid element on the freeze-out hypersurface converts into the particle distribution by use of the Cooper-Frye formula. Then the particle distribution is used to simulate the next step, which is the hadronic gas phase. Indeed, it is done by use of the ultra-relativistic quantum molecular dynamics (UrQMD) transport model \cite{Bass:1998ca}. The evolution goes on until no interaction exists in the medium and no unstable hadrons remain to decay.

We study Pb-Pb collisions with center of mass energy $\sqrt{s}=2.76$~TeV. We divide the centralities between $0$ and $80\%$ into 16 equal bins and for each bin we generate 14\,000 events. 
In the MC-Glauber, we set the wounded nucleon/binary collision mixing parameter to be $0.118$ and in the hydrodynamic evolution we choose the shear viscosity over entropy density, $\eta/s$, to be $0.08$. In this simulation, the reaction plane angle $\phi_{\text{RP}}$ has been taken to be equal to zero for all events.\footnote{ The MC-Glauber simulation data which are used in Sec.~\ref{initialSection} and \ref{cumulSec} are exactly the same data we use in the present section to study their hydrodynamic evolution. }

After generating the heavy-ion collision events, we can find the distribution of $p(v_{n,x},v_{n,y})$ in each centrality bin and consequently determine the standardized cumulants $\hat{\V}_{pq}^{(n)}$.  The results are plotted in black dots in Fig.~\ref{NonLinearResponse}. In this figure, the red dots are $\hat{\E}_{pq}^{(n)}$, similar to those in Fig.~\ref{n2cumulants} and Fig.~\ref{n3cumulants}. Recall from \eqref{linearApprox} that in the hydrodynamic linear response approximation, we have  $\hat{\E}_{pq}^{(n)}\simeq \hat{\V}_{pq}^{(n)}$. However, as can be seen from the plots in Fig.~\ref{NonLinearResponse}, $\hat{\V}_{pq}^{(n)}$ and $\hat{\E}_{pq}^{(n)}$ are not exactly the same. In fact, they have more agreements with each other in lower centralities. In higher centralities $\hat{\E}_{pq}^{(n)}$ deviates from $\hat{\V}_{pq}^{(n)}$ significantly. This means that the relation \eqref{linear} is not exact and so the  nonlinear response of hydrodynamic is important in higher centralities.

\begin{figure}[t!]
	\begin{center}
		\begin{tabular}{c}
			\includegraphics[scale=0.4]{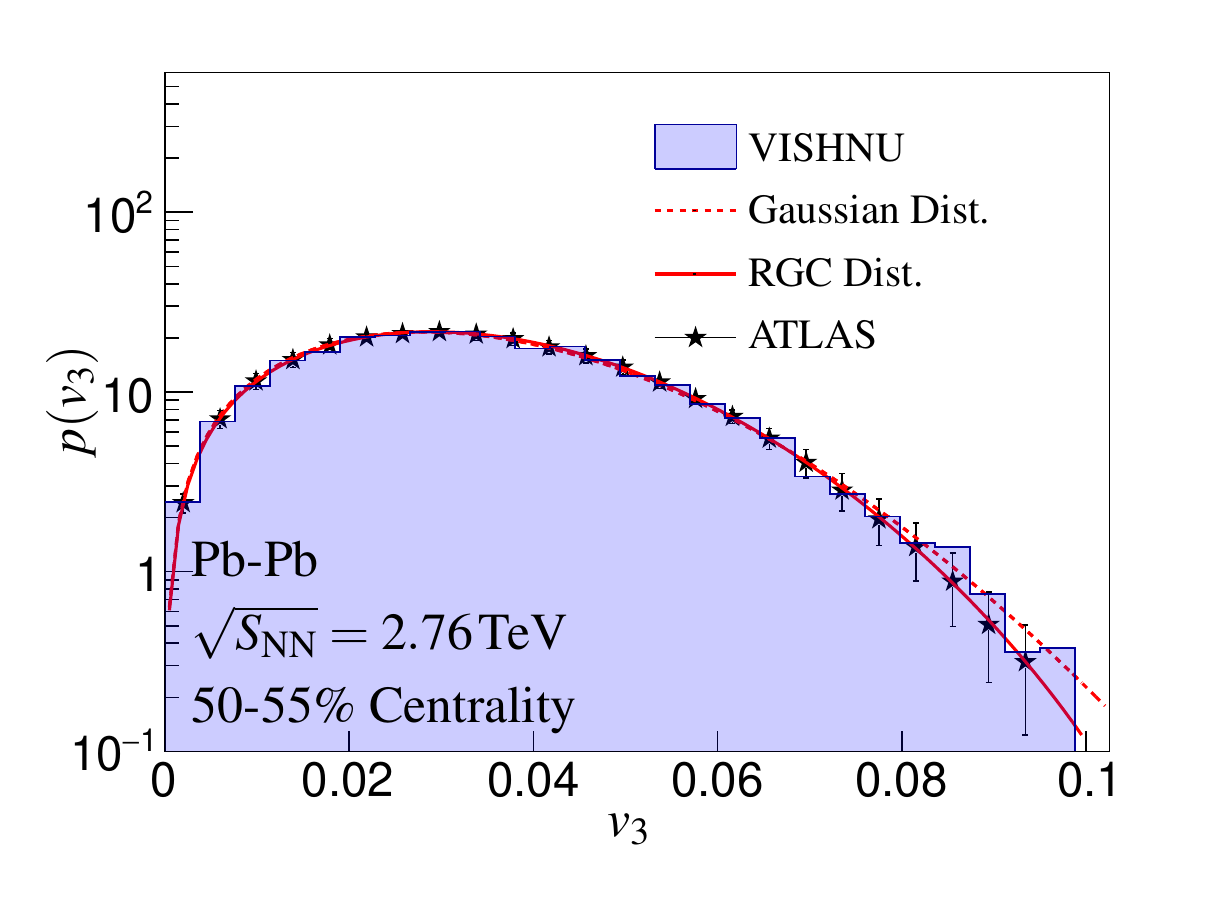}					
		\end{tabular}
		\caption{(Color online) Comparing the experimental data of ATLAS for $p(v_3)$ with MC-Glauber, Gaussian distribution, and RGC distribution.  The $\chi^2$ for Gaussian distribution fit is $\sim  3.36$ and for RGC distribution is $\sim 0.17$. The ATLAS results have been obtained from $7$-$ \mu\text{b}^{-1} $ data \cite{Aad:2013xma}.} 
		\label{RGCFitting}
	\end{center}
\end{figure}

Concentrating on the third harmonics, we use the $p(v_3)$ probability distribution reported by the ATLAS collaboration in Ref.~\cite{Aad:2013xma}. This helps us to find $\Gamma_2$ (and  $\Gamma_4$) by fitting the RGC distribution  to the ATLAS results. In Fig.~\ref{RGCFitting}, the ATLAS experimental data for $p(v_3)$ distribution is plotted in black stars for $50-55\%$ centrality.  Both Gaussian distribution (red dashed curve) and a RGC distribution  \eqref{aDistribution} (red solid curve) are fitted to the ATLAS data. As one expects, the Monte Carlo simulation has a good agreement with data. More importantly, the result obtained from the RGC distribution indicates a better fit with that of obtained from Gaussian distribution.  Note that one can fit the power distribution to data accurately as well \cite{Yan:2014afa}. However, we can find $\Gamma_2$ and $\Gamma_4$ from the RGC distribution fit unlike power distribution.\footnote{Due to the small numerical factor $\frac{1}{36}$ in \eqref{Q6}, the effect of the $\Gamma_4$ on the distribution is small and therefore we need more precise distribution to find a reasonable value via fitting. We checked that by setting $\Gamma_4=0$, the result obtained for $\Gamma_2$ is not changed drastically.} One should note that if we find the values for $v_3\{2\}$, $v_3\{4\}$, and $v_3\{6\}$ with reasonable precision experimentally, then we are able to calculate $\Gamma_2$ and $\Gamma_4$ accurately from Eqs.~\eqref{expKurt} and \eqref{expSixOrder} without fitting RGC to $p(v_3)$ distribution.

\begin{figure}[ht!]
	\begin{center}
		\begin{tabular}{c}
			\includegraphics[scale=0.4]{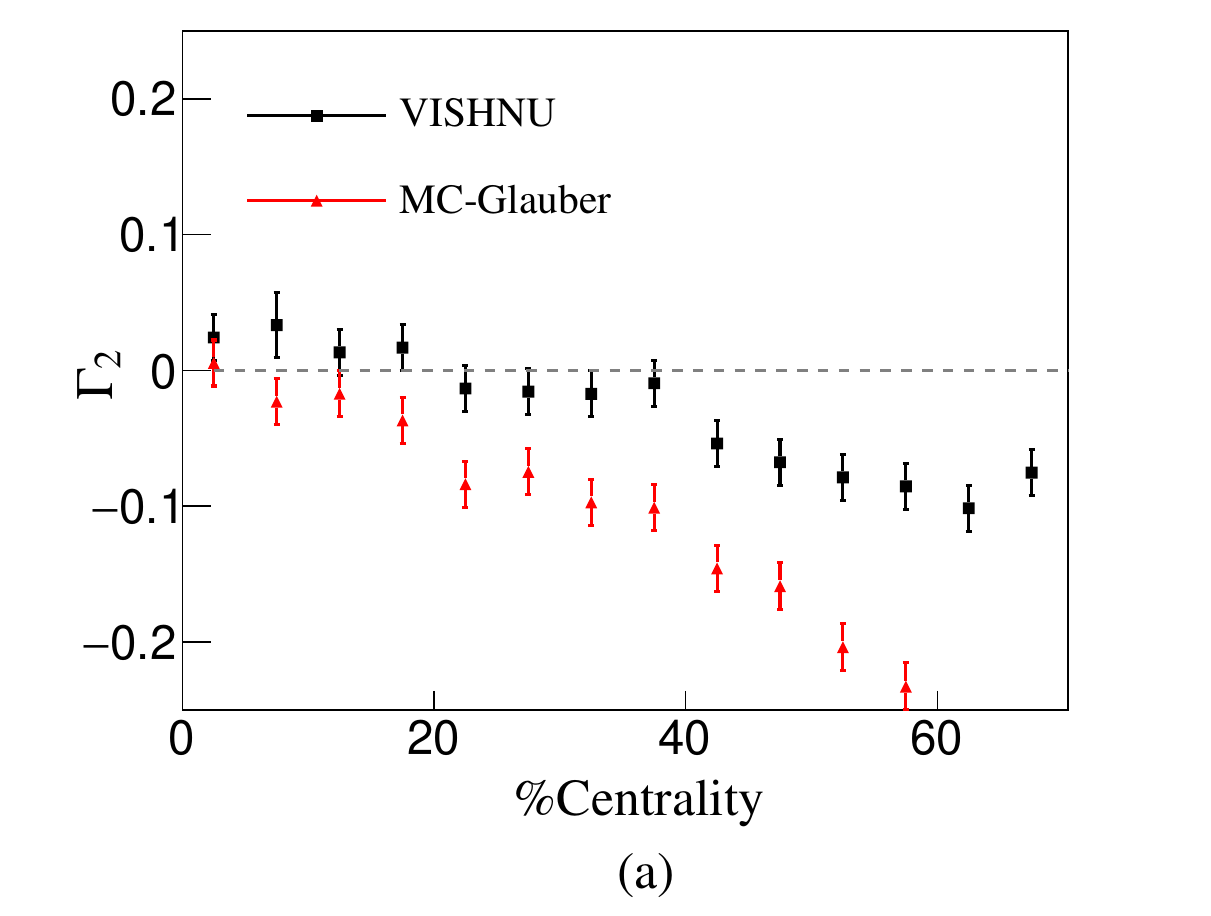}\\
			\includegraphics[scale=0.4]{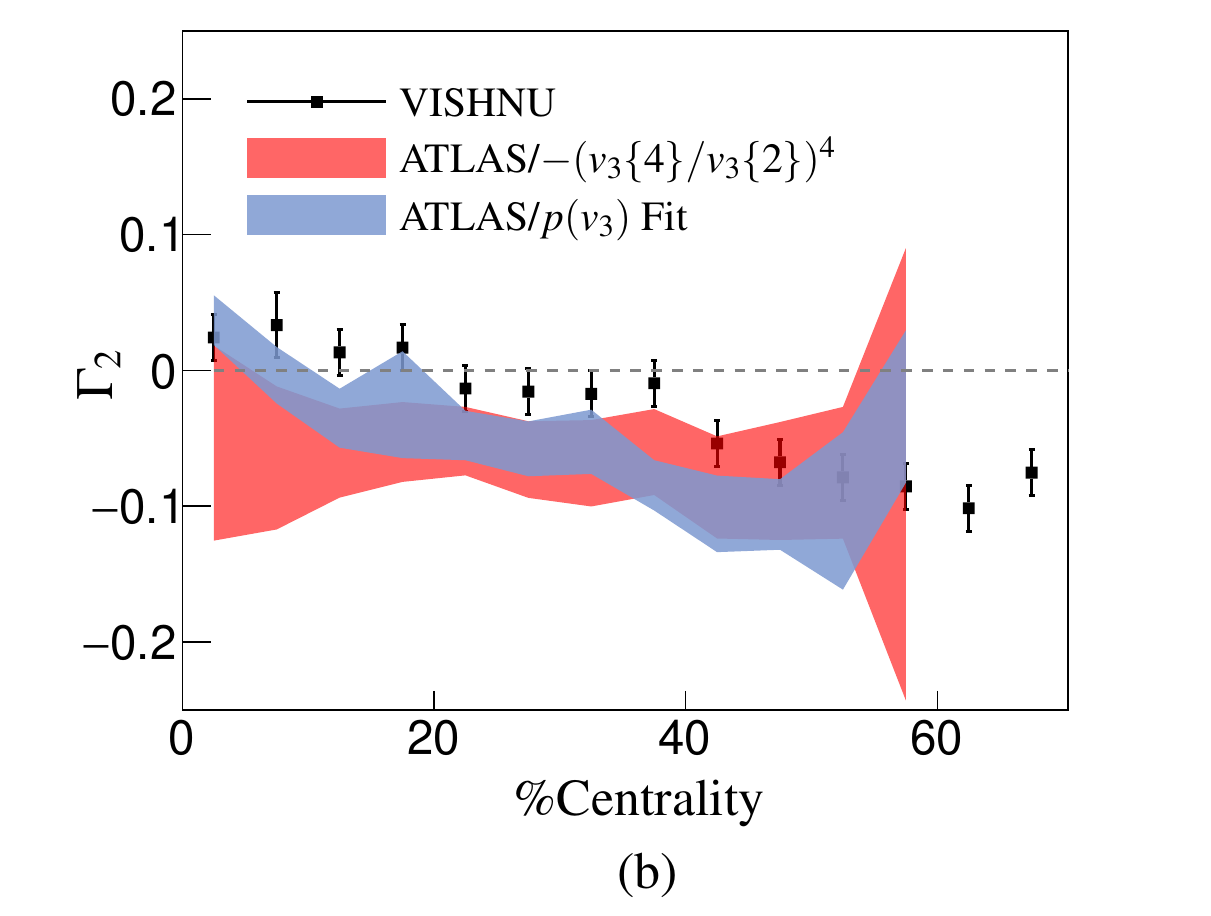}\\
			\includegraphics[scale=0.4]{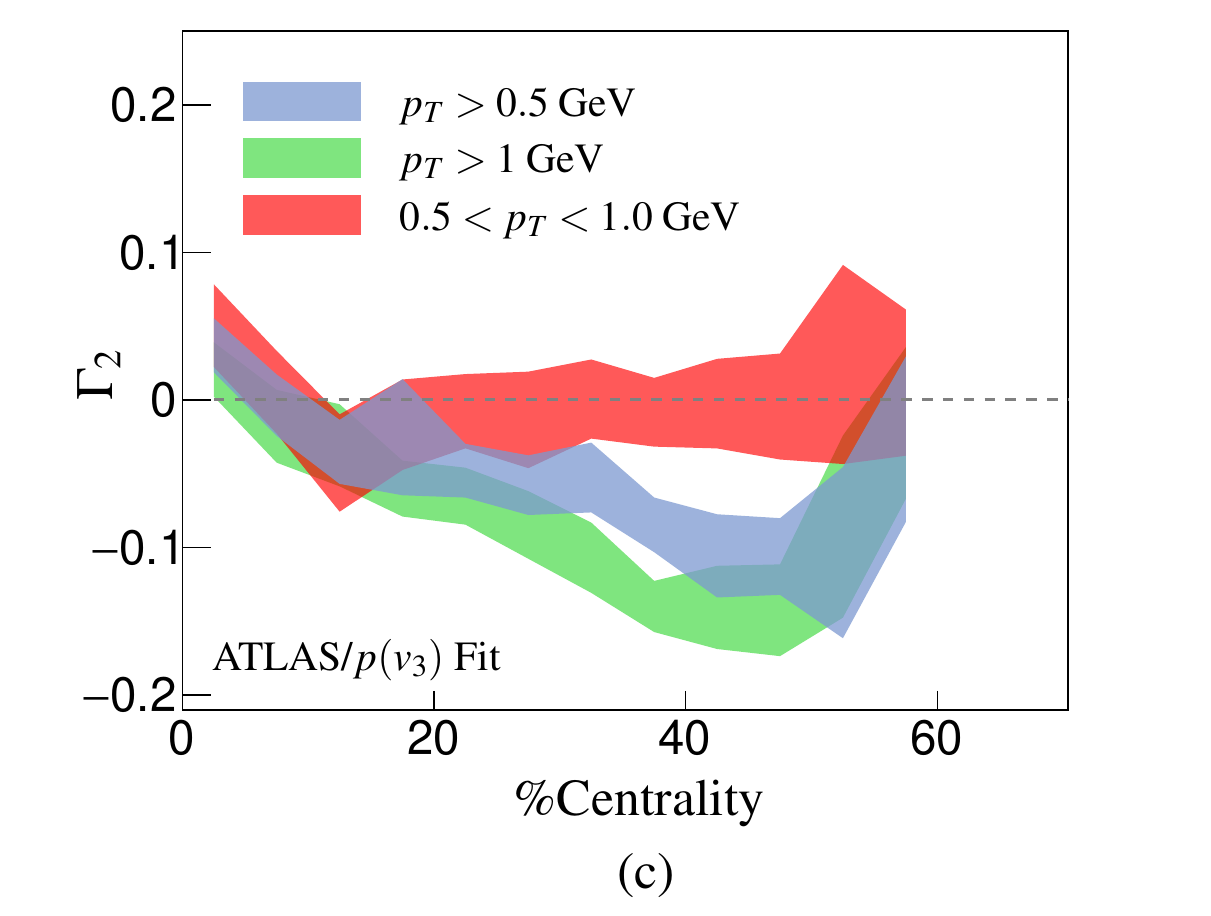}		
		\end{tabular}
	\end{center}
	\caption{(Color online)  (a) The kurtosis with respect to centrality. The kurtosis of $p(\ve_{3,x},\ve_{3,y})$ (MC-Glauber model) and corresponding distribution after the hydrodynamic evolution. (b) The kurtosis from two different methods and two different experimental data sets \cite{Aad:2014vba} and \cite{Aad:2013xma} together with VISHNU results. (c) The kurtosis in three different $p_T$ ranges as a result of fitting RGC to ATLAS data for $p(v_3)$ \cite{Aad:2013xma}. 
		\vspace*{0.3cm}
	}			
	\label{kurtosis}					
\end{figure}

From simulation, the kurtosis of $\boldsymbol{\ve}_3$ and $\boldsymbol{v}_3$ distributions can be obtained from \eqref{36}.\footnote{ For $\boldsymbol{\ve}_3$ distribution, we should replace $\hat{\V}_{pq}^{(3)}$ with $\hat{\E}_{pq}^{(3)}$ in \eqref{36}.} The results are plotted with red and black dots, respectively, in Fig.~\ref{kurtosis}(a). The initial distribution has a significant negative kurtosis; however, due to the nonlinear hydrodynamic response, the  flow distribution has a positive sign in more central collisions.

Furthermore, we can obtain the kurtosis from the experiment by using two different approaches: first, by fitting RGC to $p(v_3)$ distribution reported by ATLAS in Ref.~\cite{Aad:2013xma} and, second, by computing it directly from \eqref{expKurt} and using $v_3\{2\}$ and $v_3\{4\}$ reported by ATLAS in a separated analysis \cite{Aad:2014vba}. The $p(v_3)$ distribution in Ref.~\cite{Aad:2013xma} is reported in the three different transverse momentum windows, $p_T>0.5\;\text{GeV}$, $p_T>1\;\text{GeV}$, and $0.5<p_T<1\;\text{GeV}$. In Ref.~\cite{Aad:2014vba}, the reported transverse-momentum window for $v_3\{2\}$ and $v_3\{4\}$ is $0.5<p_T<20\;\text{GeV}$. The result is plotted in Fig.~\ref{kurtosis}(b). The red shaded region is the kurtosis calculated directly from Eq.~\eqref{expKurt} and the shaded blue region is $\Gamma_2$ obtained by fitting RGC to $p(v_3)$ with transverse momentum in the range $p_T>0.5\;\text{GeV}$. As can be seen, except in the most central collisions, there is a negative kurtosis. Also the results obtained from two different methods are in a good agreement.  This is a confirmation that $\Gamma_2$ as defined in \eqref{expKurt} contributes to the deviation of $p(v_3)$ from Gaussianity [see \eqref{aDistribution}]. 

According to Fig.~\ref{kurtosis}(b), the kurtosis predicted from iEBE-VISHNU (black dots) and that is obtained from the experimental data are almost compatible within the error bar. However, except for the most central bin, the black dots (kurtosis from iEBE-VISHNU) are slightly smaller than those obtained from the experiment.  The reason might be due to the different $p_T$ range of the experimental data and the  iEBE-VISHNU output. We have to note that the $p_T$ range of the iEBE-VISHNU outcome is $p_T\lesssim 4\;\text{GeV}$ (in fact, we used the cut-off $0.5<p_T<4\;\text{GeV}$ in our calculations), the range where the hydrodynamic works well, while the $p_T$ range of the reported data in Ref.~\cite{Aad:2014vba} (corresponds to the red shaded region) and Ref.~\cite{Aad:2013xma} (blue shaded region ) are  $0.5<p_T<20\;\text{GeV}$ and $0.5<p_T\;\text{GeV}$, respectively. It is worth mentioning that the number of particles with $p_T>4~$GeV is negligible compared to the particles in the range $0.5<p_T<4\;\text{GeV}$, and therefore, we expect a small impact on the kurtosis with particles with transverse momentum larger than $4~$GeV. However, the presence of particles with $p_T>4~$GeV in the data complicates the comparison of the data with our hydrodynamic calculations.   

 In order to study the sensitivity of the kurtosis on the $p_T$  range,  we use $p(v_3)$ distribution reported in Ref.~\cite{Aad:2013xma} for three different $p_T$ windows. The results are plotted in Fig.~\ref{kurtosis}(c) where $\Gamma_2$ has been obtained by fitting RGC to $p(v_3)$. As can be seen from the plot, the kurtosis is sensitive to the $p_T$ range. The red shaded region indicates that the distribution for softer particles with $0.5<p_T<1\;\text{GeV}$  is compatible with zero, while the distribution for more hard particles with $p_T>1\;\text{GeV}$ [green shaded region in Fig.~\ref{kurtosis}(c)] has larger kurtosis.  This is due to the fact that most of the flow is carried by the particles with transverse momentum around $3~$GeV.

 Based on the arguments above, one might deduce that the comparison between our simulation and experimental data is not well-grounded enough. To the best of our knowledge, in the center-of-mass energy $2.76\;\text{TeV}$, no $p(v_3)$ in the range of $p_T\lesssim 4\;\text{GeV}$ has been reported so far.   We expect that an experimental analysis for finding $\Gamma_2$ of $p(v_3)$ in an appropriate range of $p_T$ (for instance, $0.5<p_T<4.0\;\text{GeV}$) would lead to a more accurate compatibility. Let us note that a recent data analysis by the CMS collaboration shows a good agreement between the skewness predicted in Ref.~\cite{Giacalone:2016eyu} and the experimental data in the range $0.3<p_T<3.0\;\text{GeV}$ and with $5.02\;\text{TeV}$ center-of-mass energy \cite{Sirunyan:2017fts}.

\section{Conclusion}

In this work, we have studied the standardized cumulants of $v_2$ and $v_3$ distributions. We have modeled the ellipticity and power parameters of the elliptic-power distribution by employing the MC-Glauber model. Using this semianalytical model
together with the hydrodynamic linear response approximation, we have found that finding two-dimensional cumulants in terms of $c_2\{2q\}$ is limited to the skewness for the second harmonic. However, for the third harmonic, the higher-order  standardized cumulants can be  found  in the experiment. Specifically, the nonzero kurtosis and sixth-order standardized cumulant are responsible for nonzero values of $c_3\{4\}$ and $c_3\{6\}$, respectively.  We have found a new parametrization for the distribution $p(v_3)$ with $v_3\{2\}$, kurtosis, and sixth-order standardized cumulant being its free parameters.  It is obtained by integrating over the azimuthal angle of the two-dimensional Gram-Charlier A series. We have shown that compared to the Gaussian distribution, it suitably fits the experimental data. 

We have also compared the kurtosis obtained from experiment with that of computed by simulation. We have calculated the kurtosis from experimental data by applying two different methods: first by using $-(v_3\{4\}/v_3\{2\})^4$ and second by fitting radial Gram-Charlier distribution with $p(v_3)$ obtained from experiment. Using these methods, the quantity  $-(v_3\{4\}/v_3\{2\})^4$ shows an interesting feature. It is decreasing with centrality which is in agreement with the same quantity obtained by different initial condition models \cite{Giacalone:2017uqx}.

Here we have derived the RGC distribution for third order flow harmonic. However, it would be interesting to generalize RGC to the case of other flow harmonics. If it is fulfilled, it could be an alternative for  either elliptic-power and Bessel-Gaussian distribution \cite{Mehrabpour:2018kjs}.

\begin{acknowledgements}
     We thank Mohsen Alishahiha for encouragement and supporting the Larak-Particle-Pheno group. We also thank M. Mohammadi Najafaabdi for reading the paper thoroughly and giving useful comments. We thank to U. A. Wiedemann for discussions and useful comments on manuscript during our visit to CERN. We thank J. Ollitrault, G. Giacalone  and J. Noronha-Hostler for discussions via exchanging several emails and special thanks to J. Ollitrault and G. Giacalone for warm hospitality in the short meeting in CEA Saclay and comments on manuscript. We thank A. Akhavan for useful discussions. We thank B. Safarzadeh, M. Naseri, and H. Behnamian.  We thank participants of the ``IPM Workshop on Particle Physics Phenomenology''. We thank  the  CERN  TH  Unit  for hospitality during the final steps of this work.
     
\end{acknowledgements}

\appendix

\section{Analytical Relations for elliptic-power Moments}\label{integral}

The solution of integral \eqref{I_integral} for both even and odd values of $m$ is given by 
\begin{eqsplit}
	I_{2k}&(q,\alpha,\beta)=\sqrt{\pi}\Gamma\left(k+\frac{1}{2}\right)\Gamma\left(\alpha+1\right)\\
	&\times{}_3\tilde{F}_2(k+\frac{1}{2},\frac{\beta+1}{2},\frac{\beta}{2};\frac{1}{2},\alpha+k+\frac{3}{2};\varepsilon_0^2),
\end{eqsplit}
\begin{eqsplit}
	&I_{2k+1}(q,\alpha,\beta)=\frac{\varepsilon_0\beta\sqrt{\pi}}{2}\Gamma\left(k+\frac{3}{2}\right)\Gamma\left(\alpha+1\right)\\
	&\times{}_3\tilde{F}_2(k+\frac{3}{2},\frac{\beta+1}{2},\frac{\beta+2}{2};\frac{3}{2},\alpha+k+\frac{5}{2};\varepsilon_0^2),
\end{eqsplit}
where ${}_3\tilde{F}_2$ is the regularized hypergeometric function. Specifically,
\begin{subequations}
	\begin{eqnarray}
I_{2k}(0,\alpha-1,0)&=&\frac{\Gamma(\alpha)\Gamma(k+1/2)}{\Gamma(\alpha+k+1/2)},\\
I_{2k+1}(0,\alpha-1,0)&=&0.
	\end{eqnarray}
\end{subequations}
Using these relations, the moments of the elliptic-power distribution can be found as follows:
\bea
&&\la\ve^{k}_{n,x}\ve^{2l+1}_{n,y}\ra=0,\label{b4}\\
&&\la\ve^{2k}_{n,x}\ve^{2l}_{n,y}\ra=\nn\\
&&(X_{2k})\;{}_{3}\tilde{F}_2(k+\frac{1}{2},\alpha+1,\alpha+\frac{1}{2};\frac{1}{2},1+l+k+\alpha;\varepsilon_0^2),\nn\\\label{b5} \\
&&\la\ve^{2k+1}_{n,x}\ve^{2l}_{n,y}\ra=\nn\\
&&\ve_0(X_{2k+1})\;{}_{3}\tilde{F}_2(k+\frac{3}{2},\alpha+1,\alpha+\frac{3}{2};\frac{3}{2},2+l+k+\alpha;\varepsilon_0^2),\nn\label{b6}\\	
\eea
where
\bea
X_{2k}&=&\frac{\alpha}{\sqrt{\pi}}\left(1-\ve_0^2\right)^{\alpha+\frac{1}{2}}\Gamma(\alpha)\Gamma(k+1/2)\Gamma(l+1/2),\nn\\ \\
X_{2k+1}&=&\frac{\alpha(1+2\alpha)}{2\sqrt{\pi}}\left(1-\ve_0^2\right)^{\alpha+\frac{1}{2}}\\
&&\hspace{2.2cm}\times\Gamma(\alpha)\Gamma(k+1/2)\Gamma(l+1/2).\nn 
\eea
Note that for the case $\ve_0=0$, the only nonzero moments are as  $\la\ve^{2k}_{n,x}\ve^{2l}_{n,y}\ra$.

\section{Radial-Gram-Charlier Distribution}\label{RGCD}

\subsection{2D Gram-Charlier A Series}

The expansion of a one-dimensional distribution in terms of its cumulants  is well known (see, for instance, Ref.~\cite{kendallBook}). In this Appendix, we review the generalization of such a distribution to the two dimensions. Let us start with \eqref{cumulGen} and consider $\lambda_x\to i \lambda_x$ and $\lambda_y\to i \lambda_y$. So we can write Eq.~\eqref{cumulGen} as follows:
\begin{eqsplit}\label{generalDistCumul}
	\int d\xi_x d\xi_y \PP(\xi_x,\xi_y)&e^{i(\lambda_x \xi_x+\lambda_y \xi_y)}=\\
	\PP(\lambda_x,\lambda_y)&=\exp\left[\sum_{m,n=0}\frac{(i\,\lambda_x)^m (i\,\lambda_y)^n}{m! n!}\A_{mn}\right].
\end{eqsplit}
Note that by $ \PP(\lambda_x,\lambda_y)$ in  the second line, we mean the Fourier transformation of $\PP(\xi_x,\xi_y)$. For the special case where $\PP(\xi_x,\xi_y)$ is the 2D normal distribution
\begin{eqsplit}\label{NormalDist}
	N(\xi_x,\xi_y)=\frac{1}{2\pi \sigma_x \sigma_y}&e^{-\frac{(\xi_x-\mu_x)^2}{2\sigma_x^2}-\frac{(\xi_y-\mu_y)^2}{2\sigma_y^2}}
\end{eqsplit}
we have
\bea\label{normallDistCumul}
N(\lambda_x,\lambda_y)=\exp\left[\sum_{m,n=0}\frac{(i\,\lambda_x)^m (i\,\lambda_y)^n}{m! n!}\mathcal{N}_{mn}\right]
\eea
with the only nonzero cumulants being $\mathcal{N}_{10}=\mu_x$, $\mathcal{N}_{01}=\mu_y$, $\mathcal{N}_{20}=\sigma_x^2$, and $\mathcal{N}_{02}=\sigma_y^2$. Let us consider that the first cumulants of the distribution $\PP(\xi_x,\xi_y)$ are $\A_{10}=\mathcal{N}_{10}$, $\A_{01}=\mathcal{N}_{01}$, $\A_{20}=\mathcal{N}_{20}$, and $\A_{02}=\mathcal{N}_{02}$. By combining Eq.~\eqref{generalDistCumul} and \eqref{normallDistCumul} with each other, we can write a general distribution as
\begin{eqsplit}
	\PP(&\lambda_x,\lambda_y)=\\
	&\exp\left[\sum_{\substack{m=n=1,\\ m+n\geq 3}}\frac{(i\,\lambda_x)^m (i\,\lambda_y)^n}{m! n!}\A_{mn}\right]N(\lambda_x,\lambda_y).
\end{eqsplit}
with its  Fourier transformed being as
\begin{eqsplit}\label{GC01}
	\PP(&\xi_x,\xi_y)=\\
	&\exp\left[\sum_{\substack{m=n=1,\\ m+n\geq 3}}\frac{(-1)^{m+n}}{m! n!}\A_{mn}\frac{\partial^{m+n}}{\partial \xi_x^n\partial \xi_y^m}\right]N(\xi_x,\xi_y).
\end{eqsplit}
In order to compute the derivatives in the exponential in this equation, let us note the  Hermite polynomial defined through
\bea\label{hermitDefin}
(-1)^m\frac{\partial^{m}}{\partial \xi_x^m} e^{-\frac{\xi_x^2}{2}}=He_m(\xi_x)e^{-\frac{\xi_x^2}{2}}.
\eea
Using \eqref{NormalDist}, we immediately find
\begin{eqsplit}\label{b7}
	(-1&)^{m+n}\frac{\partial^{m+n}}{\partial \xi_x^n\partial \xi_y^m} N(\xi_x,\xi_y)=\\
	&\frac{1}{\sigma_x^m\sigma_y^n}He_m(\frac{\xi_x-\mu_x}{\sigma_x})He_n(\frac{\xi_y-\mu_y}{\sigma_y}) N(\xi_x,\xi_y).
\end{eqsplit}
 Now by considering the small deviation from Gaussian, we can expand the right hand side of \eqref{GC01} in terms of number of derivatives. Then the result is
\bea
\PP(\xi_x,\xi_y)\simeq\frac{1+\mathcal{H}}{2\pi \sqrt{\A_{20}\A_{02}}} e^{-\frac{(\xi_x-\A_{10})^2}{2\A_{20}}-\frac{(\xi_y-\A_{01})^2}{2\A_{02}}}.
\eea
where
\bea
\hspace*{-0.5cm}\mathcal{H}=\sum_{\substack{m=n=1,\\ m+n\geq 3}}   \frac{h_{mn}}{m!n!} He_n(\frac{\xi_x-\A_{10}}{\sqrt{\A_{20}}})He_m(\frac{\xi_y-\A_{01}}{\sqrt{\A_{02}}}).
\eea
The coefficient $h_{mn}$ for $m+n\leq 5$ is $h_{mn}=\hat{\A}_{mn}$. For the case $m+n > 5$ the coefficient $h_{mn}$ has more complicated form. For example, if $m+n=6$, then we have
\begin{equation}\label{hmnSix}
\begin{aligned}
h_{60}&=\hat{\A}_{60}+10\hat{\A}_{30}^2\\
h_{51}&=\hat{\A}_{51}+10\hat{\A}_{30}\hat{\A}_{21}\\
h_{42}&=\hat{\A}_{42}+4\hat{\A}_{30}\hat{\A}_{12}+6\hat{\A}_{21}^2\\
h_{42}&=\hat{\A}_{42}+4\hat{\A}_{30}\hat{\A}_{12}+6\hat{\A}_{21}^2\\
h_{33}&=\hat{\A}_{33}+\hat{\A}_{30}\hat{\A}_{03}+9\hat{\A}_{12}\hat{\A}_{21}\\
h_{24}&=\hat{\A}_{24}+4\hat{\A}_{03}\hat{\A}_{21}+6\hat{\A}_{12}^2\\
h_{15}&=\hat{\A}_{15}+10\hat{\A}_{03}\hat{\A}_{12}\\
h_{06}&=\hat{\A}_{06}+10\hat{\A}_{03}^2.\\
\end{aligned}
\end{equation}
This coefficient for $m+n>6$ has a similar form. For rotationally symmetric distributions, $\hat{\A}_{mn}$ vanishes for odd $m+n$. In this case, $h_{mn}=\hat{\A}_{mn}$ for $m+n=6$, too. However, it is not true for higher values of $m+n$.

\subsection{Integration over azimuthal angle}

In this Appendix we consider a generic 2D rotationally (with respect to origin) symmetric distribution and integrate over the azimuthal angle.

Let us first change the variables $(\xi_x,\xi_y)$ to $(\xi_r,\xi_{\phi})$ with $\xi_r=\sqrt{\xi_x^2+\xi_y^2}$ and $\xi_{\phi}=\text{atan2}(\xi_y/\xi_x)$. In this special case, we have $\mu_x=\mu_y=0$ and $\sigma\equiv \sigma_x=\sigma_y$. By using the definition of Hermite polynomials we find
\bea\label{b8}
He_n(\xi_x)He_m(\xi_y)=(-1)^{m+n}e^{\frac{\xi_r^2}{2}}\frac{\partial^{m+n}}{\partial \xi_x^n\partial \xi_y^m} e^{-\frac{\xi_r^2}{2}}.
\eea
On the other hand, the multidifferentiation of an arbitrary function $f(\xi_r)$ with respect to $(\xi_x,\xi_y)$ has the following form in the polar coordinate
\bea\label{b9}
\frac{\partial^{m+n}}{\partial \xi_x^n\partial \xi_y^m} f(\xi_r)= \left(\sum_{i=1}^{m+n}g_i^{mn}(\xi_{\phi})\frac{d^i}{d\xi_r^i}\right)f(\xi_r),
\eea
with some of the coefficient functions being
\bea
g_1^{20}&=&\frac{\sin^2(\xi_{\phi})}{r},\hspace{1.4cm} g_2^{02}=\cos^2(\xi_{\phi}),\\
g_1^{11}&=&-\frac{\sin(\xi_{\phi})\cos(\xi_{\phi})}{r},\hspace{0.1cm} g_2^{11}=\sin(\xi_{\phi})\cos(\xi_{\phi}),\\
g_1^{02}&=&\frac{\cos^2(\xi_{\phi})}{r},\hspace{1.4cm} g_2^{02}=\sin^2(\xi_{\phi}).
\eea
Using \eqref{b8} and \eqref{b9}, we have
\begin{eqsplit}
	He_n(\xi_x)&He_m(\xi_y)= \\
	&(-1)^{m+n} \sum_{i=1}^{m+n}(-1)^{i}g_i^{mn}(\xi_{\phi})He_i(\xi_r).
\end{eqsplit}
Let us define the following integral,
\bea
J_{mn}(\frac{\xi_r}{\sigma})=\int_{0}^{2\pi}d\xi_{\phi} He_n(\frac{\xi_x}{\sigma}) He_m(\frac{\xi_y}{\sigma}).
\eea
The few first terms of $J_{mn}(\xi_r)$ are listed as follows:
\bea
J_{20}(\xi_r)&=&\pi \xi_r^2-2\pi,\\
J_{40}(\xi_r)&=&\frac{3 \pi  }{4 }\xi_r^4-6 \pi  \xi_r^2+6 \pi,\label{b16}\\
J_{22}(\xi_r)&=&\frac{1}{3}J_{40}(\xi_r),\label{b17}\\
J_{60}(\xi_r)&=&\frac{5 \pi  }{8 }\xi_r^6-\frac{45 \pi  }{4 }\xi_r^4+45 \pi  \xi_r^2-30 \pi,\label{b18}\\
J_{42}(\xi_r)&=&\frac{1}{5}J_{60}(\xi_r).\label{b19}
\eea
It is worth mentioning that $J_{mn}$ is nonzero only for $n=2p$ and $m=2q$. It can be also shown that $J_{mn}(\xi_r)=J_{nm}(\xi_r)$.

Consequently  the radial distribution in \eqref{gramCharlier} reads
\begin{eqsplit}
	&\int d\xi_r p(\xi_r)=
	\int \frac{\xi_r d\xi_r}{2\pi \A_2} e^{-\frac{\xi_r^2}{2\A_2}} \int d\xi_{\phi}\,(1+\mathcal{H})=\\
	& \int \frac{\xi_r d\xi_r}{2\pi \A_2} e^{-\frac{\xi_r^2}{2\A_2}}\left[2\pi+\sum_{\substack{m=n=1,\\ m+n\geq 3}}\frac{J_{mn}(\frac{\xi_r}{\sigma})(-1)^{m+n}}{m! n!}h_{mn}\right],\nn
\end{eqsplit}
where  by using Eqs.~\eqref{b16}$-$\eqref{b19} together with Eqs.~\eqref{k2}$-$\eqref{k6}, one reaches \eqref{aDistribution}.  Let us recall that $h_{mn}=\hat{\A}_{mn}$ for $m+n\leq 6$ in the rotationally symmetric distributions. In addition, we have $\A_2\equiv \A_{20}=\A_{02}$, $\A_{10}=\mu_x$, and $\A_{01}=\mu_y$ together with $\xi_r=v_3$ and $2\sigma^2=v_3^2\{2\}$  in \eqref{aDistribution}. 
\\
\\
\\
\\

\end{document}